\documentclass[11pt,a4paper]{article}
\pdfoutput=1 

\usepackage{jheppub} 
\usepackage{graphicx}
\usepackage{bm}
\usepackage[T1]{fontenc} 

\usepackage{caption}

\usepackage[utf8]{inputenc}
\usepackage[cal=cm,scr=dutchcal]{mathalfa}
\title{\boldmath Strolling along gauge theory vacua}

\author[a]{Ali Seraj,}
\author[b,c]{Dieter Van den Bleeken}
\affiliation[a]{Institute for Research in Fundamental Sciences (IPM),\\P.O.Box 19395-5531,Tehran, IRAN}
\affiliation[b]{Primary address\\
 Physics Department, Bo\u{g}azi\c{c}i University\\
 34342 Bebek / Istanbul, Turkey}
\affiliation[c]{Secondary address\\
Institute for Theoretical Physics, KU Leuven\\
3001 Leuven, Belgium}

\emailAdd{ali\_seraj@ipm.ir}
\emailAdd{dieter.van@boun.edu.tr}

\abstract{We consider classical, pure Yang-Mills theory in a box. We show how a set of static electric  fields that solve the theory in an adiabatic limit correspond to geodesic motion on the space of vacua, equipped with a particular Riemannian metric that we identify. The vacua are generated by spontaneously broken global gauge symmetries, leading to an infinite number of conserved momenta of the geodesic motion. We show that these correspond to the soft multipole charges of Yang-Mills theory. }

\def\l{{\lambda}}

\def\calg{{\mathscr G}}
\def\calk{{\cal K}}
\def\calsk{\boldsymbol{\mathscr k}}

\def\calsg{{\boldsymbol{\mathscr g}}}

\def\calm{{\mathcal{M}}}
\def\calsm{{\boldsymbol{\mathscr m}}}

\def\calo{{\cal O}}
\def\cals{{\cal S}}
\def\calss{{\boldsymbol{\mathscr s}}}
\def\calv{{\cal V}}

\def\ua{{\underline{a}}}
\def\ub{{\underline{b}}}
\def\uc{{\underline{c}}}
\def\ud{{\underline{d}}}

\def\ulm{{\underline{\ell m}}}
\def\ulmp{{\underline{\ell'm'}}}
\def\uI{{\underline{I}}}
\def\uJ{{\underline{J}}}

\def\mn{{\mu\nu}}
\def\pd{\partial}
\def\de{\delta}
\def\rg{{\mathrm{g}}}

\def\rd{{\mathrm{d}}}

\def\nat{{\mathrm{nat}}}
\def\Ad{{\mathrm{Ad}}}

\def\boundeq{{\stackrel{\ {}_{\partial M_{\,{}_{}}}}{=}}}

\def\Tr{\mathrm{Tr}}
\def\tr{\mathrm{tr}}

\def\Dperp{{\mathbb{D}}}

\def\bE{{\boldsymbol E}}

\def\grad{{\partial}}

\newcommand{\bomega}{{\boldsymbol \omega}}
\newcommand{\bTheta}{{\boldsymbol \Theta}}

\newcommand{\bL}{{\boldsymbol L}}

\newcommand{\bk}{{\boldsymbol k}}

\usepackage[normalem]{ulem}
\begin{document}
    \maketitle
    \flushbottom       
    
    \section{Introduction}
   
    \subsection{Motivation}
    In the last few years the symmetry structure of gauge theory and gravity in asymptotically Minkowski space has received renewed interest, with new connections to the IR structure of the theory, such as soft-theorems and memory effects, emerging. We refer to \cite{Strominger:2017zoo} for a pedagogical review and exhaustive list of references.
    One conclusion of that work is the previously overlooked/under-emphasized fact that theories with gauge symmetries, from Maxwell theory to General Relativity, have a huge vacuum degeneracy. This happens because a certain subset of the gauge symmetries turn out to act as global symmetries relating physically different states and furthermore some of them are spontaneously broken, generating a non-trivial space of vacua.  A natural question to ask is if the space of vacua has an interesting geometry? One key result of our work is a direct computation of this metric on the space of vacua. We find that in the Abelian case of Maxwell theory this metric is flat, but in the case of non-Abelian Yang-Mills it turns out to be a highly symmetric but nonetheless rather rich and intricate geometry. Let us point out that our computation is performed in the simplified setting of the classical theory in a finite volume with a boundary, which might be thought of as an IR-regulation.  
    
    Another motivation for our work originated in \cite{Seraj:2016jxi}, where the global gauge symmetries of Maxwell theory were studied.  While multipole moments, associated only to charged matter distributions, are not conserved, it was shown that they can be naturally completed into conserved ``multipole charges'', by associating an additional soft\footnote{The multipole charges carried by the electromagnetic field are analogues at spatial-infinity of the soft charges of \cite{Strominger:2013lka, Barnich:2013sxa} defined at null-infinity. See \cite{Campiglia:2017mua} for recent work that makes a connection between charges defined at spatial- and null- infinity.} charge to the electromagnetic field itself. This observation begs the question if there are situations without matter with a non-trivial multipole charge carried by the electromagnetic field alone? In our paper we identify static electric fields with this property. Furthermore, our construction shows that these solutions have the natural interpretation as slow motion on the space of vacua, discussed above.  
    
    The intuition that led us to connect these two notions - the geometry of the space of vacua and pure field configurations carrying nontrivial multipole charges- comes from the theory of non-Abelian monopoles, see e.g. \cite{Weinberg:2006rq,Weinberg:2012pjx} for a review. A 't Hooft-Polyakov magnetic monopole comes with three obvious parameters, namely the position of its center in $\mathbb{R}^3$ and a fourth that is more subtle, a global U$(1)$ phase. These parameters, and their generalizations to multi-monopoles, define a sub-space of solutions, known as monopole moduli space, on which the kinetic energy of Yang-Mills theory induces a natural Riemannian metric. This geometry is known to have a rich mathematical structure \cite{Atiyah:1988jp}, but its physical importance lies in the fact that it encodes the low velocity dynamics of the monopoles. In the adiabatic limit, dynamics can appear only through the time dependence of moduli, and hence the evolution corresponds to a geodesic motion on the moduli space, which is known as the Manton approximation \cite{Manton:1981mp}. Making the positional parameters time-dependent leads to moving monopoles with non-zero momentum, while making the global U(1) phase time-dependent leads to \textit{dyons} which carry non-zero electric charge \cite{Julia:1975ff}. In short: geodesic motion on moduli space leads to nonsingular, pure field solutions that carry non-trivial conserved charges associated to symmetries of the theory. Although there might not exist any way to determine the absolute position or U(1) phase of a monopole, the physical existence of such inequivalent positions and phases manifests itself in the possibility of changing these parameters in time, which leads to physically measurable charges. This is a very intuitive argument to {\it not} treat global gauge transformations as redundancies.
    
    Although the above considerations are quite standard, we are not aware - apart from \cite{Lechtenfeld:2015uka} (and some related work \cite{,Popov:2015wsa,Lechtenfeld:2015waa,Lechtenfeld:2016sgc})- of any work performing a similar analysis for vacuum parameters instead of monopole parameters, as we do in this paper. We uncover rich non-trivial physics in any finite volume, although we have not shown if this indeed extends to infinite volume. The finite volume theory might be of interest on its own, but our results suggest that if performed carefully a limit to infinite volume preserving much of this physics might be possible. Although our work is quite close to \cite{Lechtenfeld:2015uka}, it improves, refines and extends it in a number of ways.
    
    Before we explain our setup and results in more detail below, let us mention that we follow the nomenclature of much of the literature on Yang-Mills theory and solitons, where the term {\it global gauge transformation} is used to denote those gauge transformations that do not reduce to the identity at the boundary/infinity. In that literature there also exists the notion of {\it large gauge transformation}, which is used for those gauge transformations that have non-trivial homotopy on the sphere at the boundary/infinity. That these two notions are different is illustrated for example by a gauge transformation that takes a non-trivial but constant value at the boundary/infinity: it is global but not large in the way we just defined. In much of the recent literature on asymptotic symmetries this distinction is not made and the term large gauge transformation is used for the whole class of global gauge transformations, not only those of non-trivial homotopy. Accordingly we also use the term {\it local gauge transformation}, to refer to those transformations that go to the identity on the boundary/infinity. This term is equivalent to what other authors call small gauge transformations.

    \subsection{Summary of results}
    Let us now more precisely summarize our setup and results.
    
    We begin by partially gauge fixing Yang-Mills theory by going to the temporal gauge $A_0=0$. In this gauge the dynamical fields of the theory are spacetime dependent spatial gauge fields $A=A_i(t,x)dx^i$, of which we think as curves on the configuration space $\calm$, the set of all gauge connections on a manifold $M$, which is contained in a spatial slice of Minkowksi space and that we assume to have a boundary $\partial M$. Although our discussion is more general, in all our explicit examples we choose $M$ to be a ball of finite radius $R$. After this gauge fixing the theory is still invariant under time-independent gauge transformations, which form the group of gauge transformations $\calg$, composed of maps from $M$ into the gauge group $G$.
    
      We define vacuum gauge fields $\bar A$ as those of minimal energy, namely time-independent gauge fields with vanishing spatial curvature, $\bar F_{ij}=0$. Many of these vacuum gauge fields will be physically equivalent, but not all. We will denote the set of physically distinct equivalence classes by $\calv$, which we refer to as the space of vacua. We formally parameterize this space by coordinates $z^a$, labeling the corresponding vacua as $\bar A(x;z)$. The Manton approximation then amounts to introducing time-dependence purely through these parameters, leading to time dependent gauge fields $A(t,x)=\bar A(x;z(t))$. Inserting this ansatz into the field equations will lead to equations for the curves $z(t)$ on $\calv$, that (in the low velocity approximation) can be used to analyze $\calv$ and its geometry. 
    
    In the temporal gauge the field equations split into the Gauss constraint, which is first order in time derivatives, and some remaining dynamical equations that are second order in time derivatives. As we discuss in detail the Gauss constraint removes motion along local gauge directions, i.e. those generated by the group $\calg_0$ of gauge transformations that are trivial on the boundary $\partial M$, and implies that physical motion is only allowed in directions generated by the group of global gauge transformations $\cals=\calg_0\backslash \calg$, which is isomorphic to the group of boundary gauge transformations. Furthermore this group acts transitively on the space of vacua so that one can identify it as a homogeneous space\footnote{The quotient by the isotropy group appears to be absent in \cite{Lechtenfeld:2015uka}.}:
    \begin{equation}
    \calv\cong\cals/\calk\cong\calg_0\backslash\calg/\calk\label{firsthomdef}
    \end{equation}
    where $\calk$ is the group of isotropic gauge transformations, namely those that leave a reference vacuum $\bar A_o$ invariant. We also show that the remaining dynamical equations are equivalent to geodesic equations on $\calv$ equipped with a particular left invariant metric
    \begin{equation} 
    \bar{\rg}_{ab}(z)=\mathbb{D}_{\ua\ub}\,e^\ua_a(z)\,e^\ub_b(z)\label{firstmet}
    \end{equation}
    Here $\Dperp$ is a $z$-independent operator on the algebra of gauge transformations $\calsg$, and in the expression above appear its matrix elements expressed with respect to a basis $\lambda_\ua$ of $\calsm$, where $\calsg=\calsg_0\oplus\calsk\oplus\calsm$. Because $\calsg$, $\calsg_0$ and $\calsk$ are the algebras of $\calg$, $\calg_0$ and $\calk$ respectively, it follows via \eqref{firsthomdef} that $\calsm$ is naturally identified with the tangent space $T_o\calv$. As we explain one can choose group elements
    \begin{equation}
    g_z(x)=\exp(\lambda_\ua (x)z^a)
    \end{equation}
    to explicitly parameterize the physically inequivalent vacua and to compute the metric  \eqref{firstmet} through
    \begin{equation}
    e=e^\ua_a\,\lambda_\ua\,\rd z^a\qquad\mbox{and}\qquad e=(g_z^{-1}dg_z)_\calsm\,.
    \end{equation}
    We give full details on how all these different objects are defined and constructed in general, but let us illustrate things here with the example of 3+1 dimensional Yang-Mills theory on a spatial ball of radius $R$. In that case the basis $\lambda_\ua$ can be expressed in terms of spherical harmonics and a basis $T_I$ of the gauge algebra $\mathfrak{g}$, by splitting the index $a$ as $I$, $\ell$ and $m$:
    \begin{equation}
    \lambda_{\uI\,\ulm}=T_I\left(\frac{r}{R}\right)^\ell Y_{\ell m}\qquad \ell\geq 1\,.
    \end{equation}
    It is interesting that in this case the operator $\Dperp$ acts as the dilatation operator $\Dperp=\frac{r}{R}\partial_r$ on $\calsm$. Although precisely defined through the formulas above it remains a challenge to compute the metric \eqref{firstmet} in closed form for this example, but we show that this geometry is very non-trivial as it is a Riemannian homogeneous space more general than the more familiar symmetric or naturally reductive spaces.
    
    The next step we take is to study the geodesic problem. Because $\calv$ is a homogeneous space this is quite tractable and although we are not able to find all geodesics we determine a large class of geodesic solutions that are generated by the symmetries. Translating these geodesics back to time-dependent gauge fields we find that they correspond to static\footnote{In the non-Abelian case the electric field is not gauge invariant, with static we mean that gauge invariant quantities, such as $\Tr E^2$, are time-independent.}, purely electric, source-free fields. In the example of 3+1 dimensional Yang-Mills in a ball we find for fixed $I,\ell\geq 1, m$ and a (small) arbitrary constant $v$:
    \begin{equation}
    E=-g_z^{-1}\left(\grad \Phi\right)g_z\qquad\mbox{with}\qquad \Phi=v\, T_I\left(\frac{r}{R}\right)^\ell Y_{\ell m}\,.
    \end{equation}
    Each of these electric fields carries a particular non-vanishing conserved multipole charge that we show is identical to a conserved momentum of the geodesic problem. In the example above:
    \begin{equation}
    Q_{I\,\ell m}=P_{I\,\ell m}=\ell v R\,.
    \end{equation}
    It is important to note here the restriction $\ell \geq 1$, or in general that the constant gauge transformations are excluded. Technically this is because they find themselves inside the isotropic gauge transformations $\calk$ and are hence quotiented out in our construction. Physically this corresponds to the fact that pure gauge fields cannot carry total electric charge.
    
    In conclusion certain static electric fields have the natural interpretation of a change of vacuum in time.\\
    
    This paper is structured as follows. In section \ref{setupsec} we introduce our starting point and conventions, furthermore we collect there some definitions that we will use in the later sections. Our main results and their derivation can be found in section \ref{mainsec}. In section \ref{exampsec} we illustrate these general results in a few particular examples. We end the paper with a discussion of some open issues and possible relations to some other work in the literature in section \ref{discsec}. Some technical derivations and mathematical background have been collected in the appendices.
        
    \section{Setup, notations and conventions}\label{setupsec}
    In this section we introduce and review some of the basic ingredients for our work, simultaneously setting notation and conventions.
    \subsection{Yang-Mills theory in temporal gauge}
    We consider classical, Lorentzian Yang-Mills theory with gauge group $G$, a compact semi-simple Lie group whose algebra we will denote with $\mathfrak{g}$. For simplicity we restrict our discussion to flat 4d spacetime, but our results can be generalized to a curved background of arbitrary dimensions. The action of the theory then reads\footnote{Here $\Tr$ indicates a bi-invariant scalar product on $\mathfrak{g}$. When $\mathfrak{g}=\mathfrak{su}(N)$ it reduces to minus the standard matrix trace: $\Tr=-\tr_\mathbf{N}$.} 
    \begin{equation}\label{actionYM}
    S_\mathrm{YM}=-\frac{1}{2}\int\!d^4 x\, \Tr\, F_{\mu\nu}F^{\mu\nu}\,,
    \end{equation}
    where the gauge field $A_\mu$ appears through its curvature $F_{\mn}=2\partial_{[\mu}A_{\nu]}+[A_\mu,A_\nu]$.  
    The equations of motion are
    \begin{equation}
    D_\mu F^{\mn}=0\label{covYMeoms}\,,
    \end{equation}
    with the covariant derivative $D_\mu\cdot\equiv\partial_\mu\cdot+[A_\mu,\cdot]$. 
    
    We now partially fix the gauge freedom of the theory by imposing the temporal gauge
    \begin{equation}
    A_0=0\,.\label{tempgauge}
    \end{equation}
     In this gauge it is convenient to refine our notation by splitting the time and space coordinates as $\mu=(0,i)$, and write $x^0=t$ and $x=(x^i)$. The dynamical fields in this partially gauge-fixed theory are  the spatial components, which will be referred to as $A=A_i(t,x)dx^i$. We will denote the exterior derivative on the spatial manifold $M$ somewhat unconventionally with $\grad=dx^i\frac{\pd}{\pd x^i}$ to clearly distinguish it from the exterior derivative $\rd=\rd z^a\frac{\pd}{\pd z^a}$ on the space of vacua $\calv$, which will appear in the following sections.
    
    In temporal gauge, the equations of motion \eqref{covYMeoms} become (overdots indicate  derivatives with respect to time)
    \begin{align}
    D_i\dot A_i&=0\label{Gauss}\,,\\
    \ddot A_i&=D_jF_{ji}\,.\label{YMeom1}
    \end{align}
    The Gauss equation \eqref{Gauss} is a constraint equation, while the dynamical equations \eqref{YMeom1} are the Euler-Lagrange equations\footnote{\label{boundnote}For a consistent variational principle one should impose boundary conditions such that $\oint d\Sigma_i \Tr F^{ij}\de A_j$ vanishes, for example the Neumann boundary conditions $\left.F_{ij}\right|_{\partial M}=0$. Note that the vanishing of this boundary term also ensures the conservation of multipole charges \cite{Seraj:2016jxi}.} of a Lagrangian of \textit{natural} type\footnote{We follow the terminology of \cite{arnol2013mathematical, Stuart:2007zz} }
    \begin{equation}
    L_\nat=\frac{1}{2}\rg(\dot A,\dot A)-V(A)\,,\label{YMnat}
    \end{equation} 
    where $V(A)=\frac{1}{2}\int_M\, d^3x\, \Tr F_{ij}F_{ij}$ is the potential energy. The kinetic energy term is written via a Riemannian metric on the configuration space $\calm$ of all time-independent spatial gauge fields 
    \begin{equation}
    \rg(\delta_1 A,\delta_2 A)=\int_Md^3x\, \Tr\, \delta_1 A_i\delta_2 A_i\,. \label{natmet}
    \end{equation}
    where $\delta A\in T_A\calm$.  The time-dependent gauge fields $A(t,x)$ then correspond to curves in the infinite dimensional configuration space $\calm$. 
    
    \subsection{Various groups of gauge transformations}
    The temporal gauge \eqref{tempgauge} induces only a partial gauge fixing and the theory remains invariant under residual gauge transformations that form the group $\calg$. It is comprised of {\it time-independent} gauge transformations, i.e. the group of maps from the spatial manifold $M$ into the gauge group $G$:
    \begin{equation}
    \calg=\{g: M\rightarrow G\}\,.
    \end{equation}
    Although it is of course essential, we will often drop the pre-fix `time-independent' and simply refer to $\calg$ as the group of gauge transformations. The gauge field transforms under such gauge transformations as
    \begin{equation}
    A\mapsto g\cdot A\equiv g(x)A(t,x)g^{-1}(x)+g(x)\grad g^{-1}(x)\label{leftact}\,.
    \end{equation}
    Formally, we can think of $g\,\cdot$ as providing a left group action on $\calm$.
    The infinitesimal version of this action can be expressed through a vector field
    \begin{equation}
    \quad\delta_\gamma A=-D\gamma\,,\label{gaugetransfo}
    \end{equation}
    where $\gamma$ is an element of the Lie algebra $\calsg=\{\gamma: M \to \mathfrak{g}\}$ associated to the Lie group $\calg$.
    
    In the coming sections we will encounter a number of subgroups of $\calg$. We will collect their definitions here, to give the reader a central point to look things up.
    
    In our discussion we will assume the spatial manifold $M$ to have a boundary $\partial M$. On such a manifold one can define the group $\calg_0$ of {\it local} gauge transformations, which are all gauge transformations that reduce to the identity on the boundary:
    \begin{equation}
\calg_0\equiv\{g\in \calg\,|\, \left.g\right|_{\partial M}=1\in G\} \,.   \label{Local symmetry group}
    \end{equation}
    As this group of local gauge transformations is a normal subgroup one can define the quotient group\footnote{See appendix \ref{aphom} for some details on quotient groups and more generally homogeneous spaces.}
    \begin{equation}
    \cals\equiv\frac{\calg}{\calg_0}\,.
    \end{equation}
    We will refer to $\cals$ as the group of {\it global } gauge transformations. 
    
    Given a gauge field $A$, we define its group of {\it isotropy} gauge transformations $\calk_A$ as
    \begin{equation}
    \calk_A\equiv\{g\in \calg\,|\, g\cdot A=A\}\,.
    \end{equation}
    These various groups $\calg, \calg_0, \cals$ and $\calk_A$ have corresponding Lie-algebras which we denote respectively with $\calsg, \calsg_0, \calss$ and $\calsk_A$. In the coming sections we will furthermore discuss how one can naturally define the decompositions
    \begin{equation}
    \calsg=\calsg_0\oplus\calss\,,\qquad \calss=\calsk\oplus\calsm\,.
    \end{equation}
    We also decided to indicate elements of the various subgroups/algebras with separate notation:
    \begin{align*}
    &\gamma\in \calsg\,,\quad \eta\in\calsg_0\,,\quad \sigma\in\calss\,,\quad \kappa\in\calsk\,,\quad\lambda\in\calsm,\\
    &g\in \calg,\quad h\in \calg_0,\quad s\in \cals,\quad k\in\calk \,.
    \end{align*}

    \subsection{Vacuum gauge fields and the space of vacua}\label{YMspcl}

    A {\it vacuum gauge field} $\bar A$ is defined as a solution with absolute minimum energy. As both the kinetic and potential energies introduced in \eqref{YMnat} are manifestly positive, the vacuum configurations satisfy the properties
    \begin{equation}
    \frac{\pd}{\pd t}{\bar{A}}=0, \qquad \bar F=0\label{vaccond}\,.
    \end{equation}
    Note that these conditions imply that all vacuum gauge fields $\bar A$\   solve the equations (\ref{Gauss},\,\ref{YMeom1}). Assuming that $M$ is simply connected the conditions \eqref{vaccond} imply that vacuum gauge fields are pure gauge:
    \begin{equation}
    \bar A=g\grad g^{-1}\label{pure}\,.
    \end{equation}
    Applying a gauge transformation on a vacuum gauge field results again in a vacuum gauge field. Furthermore, one can write any vacuum gauge field as the gauge transform of a reference vacuum gauge field $\bar A_o$:
    \begin{equation}
    \bar A=g\cdot \bar A_o\,.\label{vacconfig}
    \end{equation}
    A natural choice for $\bar A_o$ is $\bar A_o=0$, which we will make in our examples.
    
    Since vacuum gauge fields are time independent, they correspond to points in the configuration space $\calm$ and the space of vacuum gauge fields is a subset of $\calm$. Although the space of vacuum gauge fields is very large, many of its elements will be physically equivalent. We will call such an equivalence class a {\it vacuum} and their set the {\it space of vacua} $\calv$. 
    
    One might wonder why not all vacuum gauge fields are physically equivalent, as they are related by a gauge transformation. However, as is quite established, one should only physically identify those fields that differ by a {\it local} gauge transformation, which means that in the presence of a boundary there is a non-trivial space of vacua.   
Indeed, without taking this as input, we will arrive at the same conclusion by carefully studying the low velocity dynamics of the theory in the following sections, in addition it will reveal additional geometric properties and physical insight.
    
    The statement \eqref{vacconfig}, that any vacuum can be written as the gauge transform of a reference vacuum, ensures that the space of vacua $\calv$, although non-trivial, will have the rather simple structure of a homogeneous space, as we will discuss in detail.

    \section{Adiabatic geodesic motion on the space of vacua}\label{mainsec}
    As is well-known from scalar field theories and the physics of solitons, at low velocities the dynamics of the field theory reduces to that of geodesic motion on the space of equipotential solutions, see e.g. \cite{,Manton:1981mp,Stuart:2007zz,Weinberg:2012pjx}. In this section we repeat this argument in the case of pure Yang-Mills theory for zero-potential solutions. Working through this procedure provides two interesting insights. First of all, being able to identify particular time dependent solutions to (\ref{Gauss},\,\ref{YMeom1}) as slow motion on the space of vacua $\calv$ provides a very physical argument to why one should {\it not} physically identify gauge fields that differ by global gauge transformations. Second, this implies these global gauge transformations are physical symmetries of Yang-Mills theory and not redundancies. The geodesic motions naturally carry conserved momenta which correspond to conserved charges associated to the global gauge symmetries, as we will work out in detail. We identify the precise left invariant metric on $\calv$ that is naturally induced from the Yang-Mills Lagrangian. Surprisingly, although homogeneous, the resulting geometry turns out to be richer than that of the symmetric or normal reductive Riemannian homogeneous spaces often encountered in physics.
    
    For the ease of the readers, we have tried to make the paper self-contained. Furthermore we have emphasized the viewpoint that basically all of the mathematical structure can be deduced by a careful analysis of the physics behind the equations (\ref{Gauss},\,\ref{YMeom1}) and the Lagrangian \eqref{YMnat}.   
    
    \subsection{The Manton approximation}
    Consider the vacuum gauge fields \eqref{vacconfig}, i.e. $\bar A=g\cdot \bar A_o$. As mentioned in section \ref{YMspcl} some of these will be physically equivalent. Let us however assume that there is a subset that are physically inequivalent. We will denote this subset as $\calv$ and locally parameterize it by coordinates $z^a$. We will denote the vacua in this subset as $\bar A(x;z)$ and the corresponding gauge transformations connecting them to the reference vacuum $\bar A_o$ as $g_z$:
    \begin{equation}
    \bar A(z)=g_z\cdot\bar A_o\,.
    \end{equation}
    
    Consider now a time dependent solution $A(t,x)$ to the equations of motion (\ref{Gauss},\,\ref{YMeom1}), such that at some instant $t=t_0$ it coincides with  one of the vacua, i.e. $A(t_0,x)=\bar A(x;z_0)$.  In case the initial velocity $\dot A(t_0)$ is tangent to $\calv$ and small enough, the dynamics is well approximated by free (geodesic) motion on $\calv$ \cite{Stuart:2007zz}. Its validity is essentially guaranteed by the Lagrangian being of natural form with a positive potential as in \eqref{YMnat}. This approximation is an example of the more general principle of adiabatic motion, known in the literature on solitons as the Manton approximation \cite{Manton:1981mp}. As the initial velocity $\dot A(t_0)$ is increased, the corresponding solution will have a rather complicated time evolution, corresponding to oscillations in directions normal to $\calv$.

\begin{figure}
    \centering
    \includegraphics[width=0.7\linewidth]{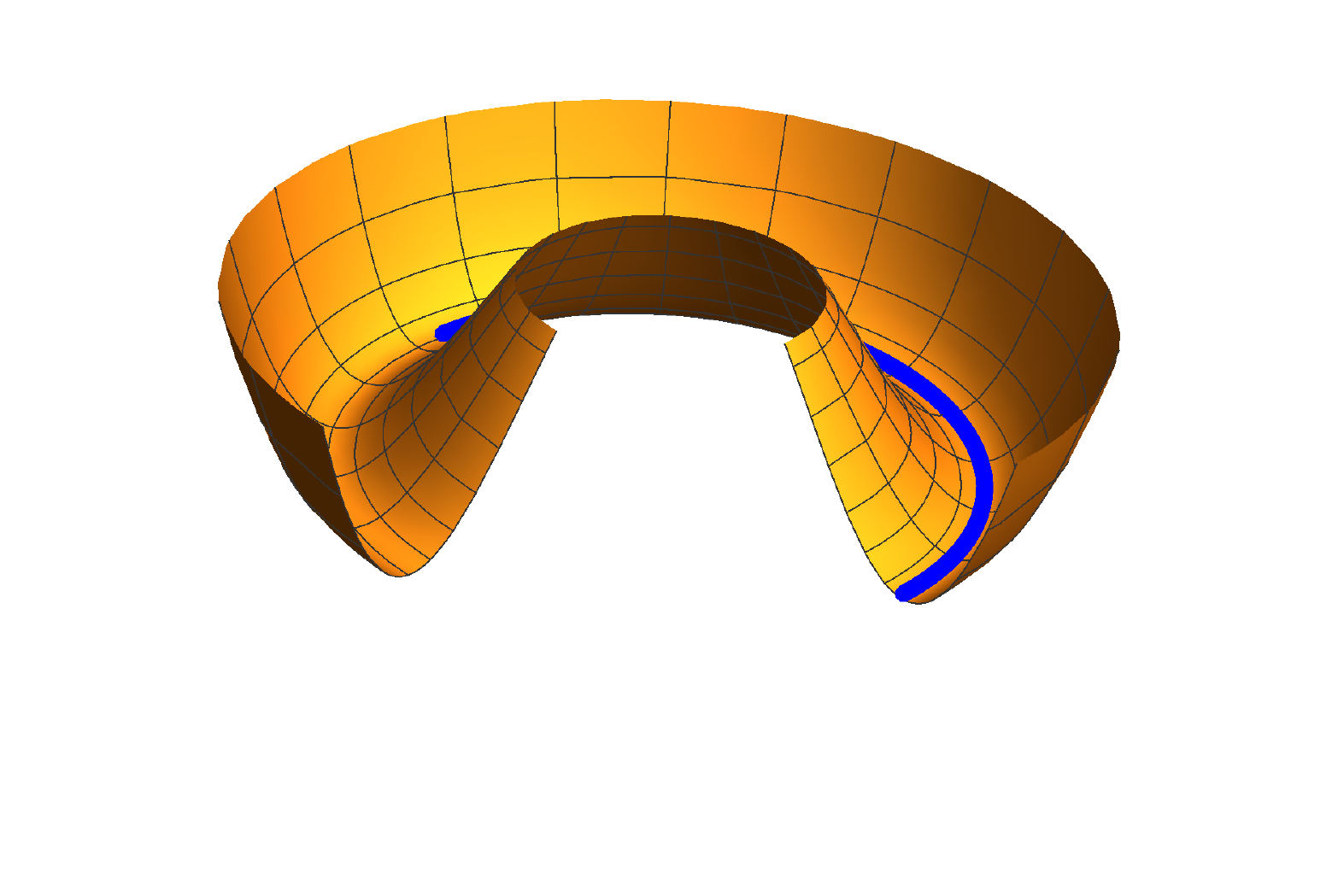}
    \captionsetup{width=0.8\textwidth}
    \vspace{-1.8cm}
    \caption{\textit{Adiabatic motion on the space of vacua}. Here a potential with a non-trivial set of absolute minima, the space of vacua, is pictured. The blue line represents slowly strolling along the vacua. By increasing the velocity the motion will start to deviate from the blue adiabatic line and move up the potential. This is a finite dimensional analog of the situation in Yang-Mills theory.}
    \label{fig:adiabatic}
\end{figure}

    In practice the approximation amounts to allowing time dependence only through the parameters $z^a$:
    \begin{equation}
    A(t,x)=\bar A(x;z(t))\,.\label{ccan}
    \end{equation}
    Such a time dependent gauge field corresponds to a curve $z(t)$ on the space of vacua $\calv$ with time the parameter along that curve. The idea is now to consider \eqref{ccan} as an ansatz to be inserted into the equations of motion, whose components along $T\calv$ will reduce to a geodesic equation, while the orthogonal components are subleading in the adiabatic limit\footnote{More precisely the adiabatic limit for a natural Lagrangian system, $L=\rg_{\alpha\beta}(\phi)\dot\phi^\alpha\dot\phi^\beta-V(\phi)$\,, can be defined by rescaling time as $\tau=\epsilon t$, expanding the fields as $\phi^\alpha(t)=\phi_{\star}^\alpha(\tau)+\delta^\perp\phi^\alpha(\tau,\epsilon)$ and sending $\epsilon\rightarrow 0$. Here $\delta^\perp\phi$ is orthogonal, w.r.t.\! the metric $\rg$, to the zero-mode directions $\delta^\parallel\phi^\alpha$, for which $\delta_\alpha\delta_\beta V\delta^\parallel\phi^\beta=0$. It follows from the equations of motion that $\phi_\star$ is constrained to the subspace $\delta_\alpha V=0$. Furthermore the non-zero-modes decouple from the equations of motion in directions parallel to the zero-modes. These reduce to the geodesic equations on the surface $\delta_\alpha V=0$, while the remaining equations imply that $\delta^\perp\phi^\alpha(t,\epsilon)=\calo(\epsilon^2)$ and hence can be ignored in the adiabatic limit.\label{adiabatic limit}}. Equivalently one can insert ansatz \eqref{ccan} into the Lagrangian to directly obtain the effective Lagrangian describing point particle motion on $\calv$, endowed with a particular metric. We will follow this second route, as the geometric structure appears more naturally, and in section \ref{seceoms} show that it is equivalent to the approach at the level of the equations of motion. In this second approach we should be careful however, as \eqref{Gauss} is a constraint that does not follow from the Lagrangian \eqref{YMnat}. Therefore, we will solve $\eqref{Gauss}$ exactly on the ansatz \eqref{ccan} and evaluate the Lagrangian \eqref{YMnat} on these solutions only. 
    
    
    By construction the potential vanishes on the ansatz \eqref{ccan}, $V(\bar{A}(z))=0$, so the Lagrangian \eqref{YMnat} reduces to the kinetic term. Via \eqref{vacconfig} and \eqref{leftact}, one observes that the ansatz \eqref{ccan} really amounts to
    \begin{align}
\nonumber    A(t,x)&=g_{z(t)}(x)\cdot\bar A_o(x)\\
    &=g_{z(t)}(x)\bar A_o(x)g_{z(t)}^{-1}(x)+g_{z(t)}(x)\grad g_{z(t)}^{-1}(x)\label{gauge fieldexpl}\,.
    \end{align}
    Let us stress that the time dependent transformations $A\to g(t,x)\cdot A$ are {\it not} symmetries of the gauge fixed theory \ref{YMnat}. In particular the time dependent gauge field above should not be confused with a pure gauge connection, as it has a non-trivial field strength.
    
    Indeed, one can now compute the electric field $E_i=F_{0i}=\dot A_i$. Geometrically it corresponds to the tangent vector of the curve on $\calv$ described by \eqref{ccan}{}, and not surprisingly it can be expressed instantaneously as an infinitesimal gauge transformation: 
    \begin{align}\label{electric field tangent}
    \dot A&=\delta_{\gamma_z}\bar A=-D\gamma_z\,, \qquad \gamma_z=\dot g_z g^{-1}_z\in \calsg\,.
    \end{align}
    A crucial feature, originating form the fact that any vacuum can be obtained by a left action on the reference vacuum $\bar A_o$, is that one can relate the tangent vector at $\bar A$ to one at the reference point, $\delta_{\gamma_z}\bar A=g_z(\delta_{\sigma_z}\bar A_o)g_z^{-1}$\,, so that :
    \begin{equation}
    \dot A=-g_z\,({D_o}\sigma_z)\,g_z^{-1}\,,\qquad \sigma_z=g_z^{-1}\dot g_z\label{lambdadef}\,.
    \end{equation}
      
    Evaluating \eqref{YMnat} on the ansatz \eqref{ccan} then leads to the effective Lagrangian:
    \begin{equation}
    L[z(t)]=\frac{1}{2}\bar{\rg}_z(\dot z,\dot z)\label{LeffYM}\,.
    \end{equation}
    Geometrically speaking $\bar{\rg}$ is the metric induced on the space of physical vacua $\calv$ by embedding it in the configuration space $\calm$, with the metric \eqref{natmet}. On vectors of the form \eqref{lambdadef} it reduces to
    \begin{equation}
    \bar{\rg}_z(\dot z ,\dot z)=\rg(D_o\sigma_z,D_o\sigma_z)\label{metric1}\,.
    \end{equation}
    The right hand side defines a ($z$-independent) inner product on the gauge algebra $\calsg$:
    \begin{equation}
    \rg(D_o\gamma_1,D_o\gamma_2)=\int_M d^3x\,\Tr\,D_{o}^i\gamma_1  D_{o}^i\gamma_2\,,\qquad \gamma_{1},\gamma_{2}\in \calsg \,.\label{scalar product YM}
    \end{equation} 
    In the next subsections we will see how this inner product further simplifies when restricted to those gauge parameters that satisfy the same physical properties as $\sigma_z$ and how it encodes the geometry all over $\calv$.
    
    \subsection{The Gauss constraint and global gauge transformations}\label{gaussglobal}
    
    Let us start by imposing the Gauss constraint equation \eqref{Gauss}. A motion on the space of physical vacua $\calv$ satisfies \eqref{lambdadef}, so that the Gauss constraint becomes 
    \begin{equation}
    D_i\dot{A}^i=-g_z (D^2_o\sigma_z)\,g_z^{-1}=0,\qquad\qquad D^2\equiv D_{i}D_{i}\label{D2}\,.
    \end{equation}
    The above equation implies that for any curve $z(t)$ on $\calv$ 
    \begin{equation}
    D^2_o\sigma_z=0 \label{YMGauss}\,.
    \end{equation}
    Let us denote with $\calss$ the vector space of solutions to this equation
    \begin{equation}
    \calss\equiv\ker D^2_o\subset\calsg\label{YMbreve}\,.
    \end{equation}
    The above equation can be thought of as a constraint on vectors tangent to $\calv$, which should not be a surprise given the well-known interpretation of the Gauss equation as a Hamiltonian constraint.

    \subsubsection{Splitting infinitesimal gauge transformations}\label{secgauss}
    The above result, that tangent vectors to physical vacua map only to the subset \eqref{YMbreve} of all infinitesimal gauge transformations can be given a more geometrical interpretation. First define the subgroup $\calg_0\subset \calg$ of \textit{local gauge transformations} as the set of all gauge transformations that are equal to the identity at the boundary $\partial M$ of the spatial manifold $M$. Denoting with $\calsg_0$ the Lie algebra of infinitesimal local gauge transformations we can then give another characterization, alternative to \eqref{YMbreve}, of $\calss$: 
    \begin{quote}
    {\it The vector space $\calss$ is the complement of $\calsg_0$ in $\calsg$ and is orthogonal to it w.r.t.\! the inner product \eqref{scalar product YM}:
    	\begin{equation}
    	\calsg=\calsg_0\oplus\calss\label{decomp}\,.
    	\end{equation}}
    \end{quote}
    Accordingly each infinitesimal gauge transformation can be uniquely split as
    \begin{equation}
    \gamma=\gamma_0+\gamma_\calss\qquad\mbox{with}\quad \gamma_0\in\calsg_0\,,\ \ \gamma_\calss\in\calss\,.
    \end{equation}
    
    \noindent This alternative way to describe $\calss$ is interesting, as it gives \eqref{YMGauss} the interpretation that physical motion only takes place in directions orthogonal to local gauge directions, which in turn indicates that all local gauge directions should be considered as redundant directions containing no physical information. 
    
    The above characterization can be easily verified. Consider the inner product of a vector $\sigma\in\calsg$ and a vector $\gamma_0\in \calsg_0$. Using the definition \eqref{scalar product YM} we have
    \begin{eqnarray}
    \nonumber    \rg(D_o\gamma_0,D_o\sigma)&=&\int_M\! d^3x\, \Tr\,D_o^i\gamma_0\,D_o^i\sigma\nonumber\\
    &=&\oint_{\partial M}\! d\Sigma_i \Tr\,\gamma_0\,D_o^{i}\sigma\label{YMpi}-\int_M\! d^3x\, \Tr\,\gamma_0\,D_o^2\sigma\label{calc}\,.
    \end{eqnarray}
    The first term on the right is vanishing as $\gamma_0$ vanishes by definition at the boundary. Now if $\sigma\in \calss$, then clearly the inner product vanishes due to \eqref{YMGauss}. Conversely, if the inner product vanishes for any $\gamma_0\in \calg_0$, then $D_o^2\sigma=0$ everywhere which implies that $\sigma\in \calss$. Note that the inner product \eqref{scalar product YM} is degenerate and so a priori there might be a non-trivial overlap between $\calsg_0$ and $\calss$. This is however not the case, as we show in appendix \ref{diffeqapp}, and $\calsg_0$ and $\calss$ are each others complement.
    
    This discussion also leads us one step further in the computation of the metric appearing in the effective Lagrangian \eqref{LeffYM}. As the $\sigma_z$ satisfy \eqref{YMGauss}, it is enough to work out the restriction of the inner product \eqref{scalar product YM} to $\calss$. Let us define this restriction through
    \begin{equation}
    \langle\sigma_1,\sigma_2\rangle\equiv\rg(D_o\sigma_{1},D_o\sigma_{2})\,,\qquad \sigma_1,\sigma_2\,\in\calss\,.
    \end{equation}
    A computation similar to \eqref{calc} implies that this restricted inner product reduces to the boundary integral
    \begin{align}
\boxed{    \langle\sigma_1,\sigma_2\rangle=\oint_{\partial M}\! d\Sigma\, \Tr\,\gamma_{1} \Dperp \gamma_{2}\label{angles}}
    \end{align}
    where the operator $\Dperp$ on the boundary is simply a normal derivative
    \begin{equation}
    \left.(\Dperp\sigma)\right|_{\partial M}=\left.(n^i D_{o\,i}\sigma)\right|_{\partial M}\label{Dperpbound}\,,
    \end{equation} 
    with $n^i$ the unit normal vector to the boundary. As we will explain in section \ref{isosec}, $\Dperp$ gets naturally extended to the bulk where, as we will see in the various examples of section \ref{exampsec}, it is a linear \textit{dilatation} operator on $\calss$.

The upshot of this discussion is that since $\sigma_z\in\calss$ it follows that we can rewrite the metric \eqref{metric1} as
    \begin{equation}
    \bar\rg_z(\dot z,\dot z)=\langle\sigma_z,\sigma_z\rangle\label{YMmetric}\,,
    \end{equation}
    where via \eqref{angles} this is determined fully by the boundary data.
    
    \subsubsection{Global symmetries as a quotient group and the modified bracket}\label{secglob}
    
    The vector space $\calss$ contains the tangent space of the space of physical vacua $\calv$ at the reference point $\bar A_o$. However, $\calss$ is not a subalgebra of $\calsg$ with respect to the natural bracket induced form $\calsg$. In this section, we present another characterization of $\calss$ that introduces a unique \textit{modified} bracket on it so that it becomes an algebra. 
    
    Let us start with the local gauge transformations, defined in \eqref{Local symmetry group}. Clearly $\calg_0$ is a subgroup of $\calg$, and furthermore a \textit{normal} subgroup.  The remaining gauge transformations, namely those not going to the identity at the boundary $\partial M$ do not immediately form a group. If one however identifies all transformations that differ by a local gauge transformation, then these equivalence classes have a natural group structure\footnote{For a brief introduction to quotient groups and homogeneous spaces, see appendix \ref{sec-homogeneous spaces}.}, i.e. the \textit{cosets} $\calg_0g$ form the \textit{quotient group} 
    \begin{equation}\label{quotient_group}
    \cals\equiv\calg_0\backslash\calg\,.
    \end{equation}
    We will refer to $\cals$ as the group of \textit{global gauge transformations}. 
    Equivalently, the Lie algebra of local gauge transformations $\calsg_0$ is a Lie algebra \textit{ideal}, which guarantees that the equivalence classes $\gamma+\calsg_0$ of infinitesimal global gauge transformations have a natural Lie algebra structure with respect to the bracket
    \begin{equation}\label{coset algebra}
    [\gamma_1+\calsg_0,\gamma_2+\calsg_0]=[\gamma_1,\gamma_2]+\calsg_0\,,
    \end{equation}
    where the Lie bracket on the right hand side is that of $\calsg$. This quotient algebra $\calsg_0\backslash \calsg$, which indeed is the Lie algebra of the group $\cals$, is formally made up of equivalence classes but can be given a more explicit representation by choosing a unique representative for each class: 
    
    \begin{quote}
    \textit{The Lie algebra of $\cals$ can be identified with the vector space $\calss$, defined in \eqref{YMbreve}, equipped with a modified bracket, introduced in \eqref{YMcom}.}
    \end{quote}
    
    Let us go through the steps that lead to the statement above. As was mentioned, the algebra of gauge transformations can be split into $\calsg=\calsg_0\oplus\calss$ and hence any infinitesimal gauge transformation $\gamma$ can be decomposed uniquely as $\gamma=\gamma_0+\gamma_\calss$. The component $\gamma_\calss$ is the unique $\sigma\in\calss$ such that $\left.\sigma\right|_{\partial M}=\left.\gamma\right|_{\partial M}$. As $\calss$ is the space of solutions of the second order differential equation $D_o^2\sigma=0$, in practice $\gamma_\calss$ is the solution of the Dirichlet problem given the boundary value of $\gamma$ whose uniqueness is shown in appendix \ref{diffeqapp}. The key point is now that $\gamma_\calss\in \calss$ can be regarded as a unique representative for each class $\gamma+\calsg_0$\,. Indeed, if $\gamma_1=\gamma_2+\calsg_0$ then $\left.\gamma_1\right|_{\partial M}=\left.\gamma_2\right|_{\partial M}$ and thus $(\gamma_1)_\calss=(\gamma_2)_\calss$. This establishes that $\calss$ and $\calsg_0\!\backslash \calsg$ are identical as vector spaces. 
    
    As $\calss$ is a subspace of $\calsg$ it has a natural bracket, which is simply the restriction of that of $\calsg$, i.e. the commutator $[\cdot,\cdot]$. However, this bracket does not close on $\calss$
    \begin{align}
    D_o^2 [\sigma_1,\sigma_2]&=2[D_o^i \sigma_1, \,D_o^i \sigma_2]\neq 0\,.
    \end{align}
    However, we can define a {\it modified bracket} $[\cdot,\cdot]_*$, that does close on $\calss$ by adding a suitable local gauge transformation\footnote{Similar modified brackets, defined to bring back the commutator to a specified form by a suitable local gauge transformation, have appeared in other settings as well \cite{Barnich:2010eb,Compere:2015knw}.}. Equivalently we can use the bijection between $\calss$ and $\calsg_0\backslash \calsg$ to induce its bracket on $\calss$. Explicitly, the modified bracket is simply the commutator projected onto $\calss$:
    \begin{equation}
    [\sigma_1,\sigma_2]_*\equiv [\sigma_1,\sigma_2]_\calss \label{YMcom}\,.
    \end{equation}

    \subsubsection{Global gauge symmetries as boundary symmetries}\label{sec-ASG}
    Since two global gauge transformations in a common equivalence class differ by a local gauge transformation, it follows that they are identical on the boundary, in particular $\left.\gamma\right|_{\partial M}=\left.\gamma_\calss\right|_{\partial M}$. This implies that on the boundary the modified bracket reduces to the usual commutator:
    \begin{align}
    \left.[\sigma_1,\sigma_2]_*\right|_{\partial M}&=\left.[\sigma_1,\sigma_2]\right|_{\partial M}\label{starbound}\,.
    \end{align}
This indicates that the group of global gauge transformations $\cals$ is naturally isomorphic to the group of boundary gauge transformations:
    \begin{equation}
    \cals\cong \left.\calg\right|_{\partial M}\,,\qquad\qquad \left.\calg\right|_{\partial M}=\{\left.g\right|_{\partial M}:\partial M\rightarrow G\}\label{bulkbound}\,.
    \end{equation}
    Indeed, this follows more formally from applying the first isomorphism theorem (discussed in appendix \ref{sec-homogeneous spaces}) to the homomorphism 
    \begin{equation}
    \left.{}\right|_{\partial M} :\calg\rightarrow  \left.\calg\right|_{\partial M} : g \mapsto \left.g\right|_{\partial M}\,.
    \end{equation} 
    Clearly $\ker \left.{}\right|_{\partial M}=\calg_0$ while $\mathrm{Im} \left.{}\right|_{\partial M}=\left.\calg\right|_{\partial M}$, so that by the theorem, $\cals=\calg_0\backslash \calg\cong \left.\calg\right|_{\partial M}$\,. It is interesting that this group of boundary gauge transformations also appears as the asymptotic symmetry group of Yang-Mills theory \cite{Strominger:2013lka, Barnich:2013sxa}.\\ 
    
    In summary there are three equivalent ways of presenting an infinitesimal global gauge transformation $\sigma \in \calss$:
    \begin{align}
    \begin{split}
    \ker D_o^2&\quad\leftrightarrow\quad\;\calsg_o\!\backslash \calsg\;\;\quad\leftrightarrow\,\quad\left.\calsg\right|_{\partial M}\,,\\
    \sigma\quad&\quad\leftrightarrow\quad\sigma+\calsg_0\quad\leftrightarrow\quad \left.{\sigma}\right|_{\partial M}\,.
    \end{split}
    \end{align} 
    From the Yang-Mills theory perspective, which is defined in the bulk $M$, the first point of view is the most straightforward one, as it allows us to explicitly represent the global gauge transformations as a particular subset of gauge transformations acting non-trivially everywhere on $M$. Conceptually on the other hand the last representation seems simpler and it is technically easier to work with boundary values only. The metric on the space of vacua can be defined purely in terms of boundary values, as we saw in \eqref{YMmetric}, and thus in much of our discussion it will suffice to consider only fields on the boundary. The bulk perspective however enables us to obtain explicit spacetime fields corresponding to adiabatic motions on the space of vacua as we will show in section \ref{sec-geodesics}.  The equivalence above implies that all results obtained on the boundary can be translated into the bulk, and we will provide the necessary details for this in section \ref{secreps}.

    \subsection{Isotropic gauge transformations and the second quotient}\label{isosec}
    We have seen how the Gauss constraint restricted physical tangent vectors to the subspace $\calss$ orthogonal to local gauge transformations. Moreover, the space $\calss$ was related to a quotient of the group of gauge transformations. In this section, we discuss a second restriction removing isotropic gauge transformations and thus introducing a second quotient, which will bring us to the final form of the space of vacua as a homogeneous space.
    
    \subsubsection{The kernel of $\Dperp$ and isotropic gauge transformations}
    Consider the bi-invariant inner product on $\calss$ that naturally follows from its equivalence to the algebra of boundary gauge transformations:
    \begin{equation}
    (\sigma_1,\sigma_2)\equiv\oint_{\partial M}\! d\Sigma \,\Tr\, \sigma_1\sigma_2\label{bndKilling}\,.
    \end{equation}
    Now note that the metric \eqref{YMmetric} on the space of vacua involves an inner product that is not simply the one above, but is related to it through the operator $\Dperp$:
    \begin{equation}
    \langle\sigma_1,\sigma_2\rangle=(\sigma_1,\Dperp \sigma_2)\label{innerproducts}\,.
    \end{equation} 
    In this section, we discuss the important role of this operator and how it affects the space of physical vacua. 
    We will show that those global gauge transformations in the kernel of this operator are exactly the isotropic symmetries leaving the reference vacuum invariant. Accordingly, the space of physical vacua corresponds to an additional quotient of the group of global gauge symmetries by this isotropy group.

    So far the action of the operator $\Dperp$  is defined in \eqref{Dperpbound} only on the boundary. However, it can be uniquely extended to the bulk of $M$ by demanding that its image is in $\calss$. That this extension is unique follows from the observation, derived in appendix \ref{diffeqapp}, that all elements of $\calss$ are uniquely determined by their boundary values:
    \begin{align}
    \begin{split}
        &\Dperp:\ \calss\to \calss:\quad \sigma\mapsto \Dperp\sigma\\
\mbox{such that}    \qquad \left.(\Dperp\sigma)\right|_{\partial M}&=\left.(n^i D_{o\,i}\sigma)\right|_{\partial M}\qquad\mbox{and}\qquad D^2_o\, \Dperp\sigma=0\label{Dperpdef}\,.
    \end{split}
    \end{align}
    Here $n$ is the unit normal vector field to $\partial M$. Note that $\Dperp$ is a symmetric operator with respect to the inner product \eqref{bndKilling}, as follows from \eqref{calc}.
    
    Now let us investigate the physical role of the operator $\Dperp$, and why it appears in the effective Lagrangian. A priori one could imagine $\sigma_z\in \calss$, associated to some motion $z(t)$, to be in the kernel of $\Dperp$ . Accordingly the effective Lagrangian \eqref{YMmetric} would vanish, implying that there would be no kinetic energy associated to such motion. Clearly that makes no sense, and the resolution is that such directions do not correspond to a change of vacuum, since those elements $\sigma\in \calss$ in the kernel of $\Dperp$ turn out to act trivially on the reference vacuum $\bar A_o$.
    
    To explain this, let us define the group of isotropic gauge transformations  as 
        \begin{align}
        \calk=\{k\in\calg: k\cdot \bar A_o=\bar A_o \}\,.
        \end{align}
    The corresponding algebra $\calsk$ is thus given by elements $\kappa\in\calss$ such that $\de_{\kappa}\bar A_o=D_o\kappa=0$, i.e $\calsk\equiv\ker D_o$. Now we can state our main observation:
\begin{center}
    \textit{The kernel of the operator $\Dperp$ coincides with the algebra $\calsk$ of isotropic gauge transformations.}
\end{center}
    The above is equivalent to\footnote{Note that $D_o$ is a map from $\calsg$ to $\mathbb{R}^3\otimes \calsg$, contrary to $\Dperp$ which maps $\calss$ to $\calss$.} $\ker \Dperp=\ker D_o$, which we show in appendix~\ref{diffeqapp}. In summary, the presence of the operator $\Dperp$ guarantees the removal of the isotropic gauge directions.

 	This suggests to define the decomposition
	\begin{equation}
	\calss=\calsk\oplus \calsm\,,\qquad \calsm=\calsk^\perp\,,\label{algebradecomp}
	\end{equation}
	with the orthogonal complement taken with respect to the metric \eqref{bndKilling}. We can then conclude from the discussion above that $\langle\cdot,\cdot\rangle$ is a positive definite inner product when restricted to $\calsm$\,, furthermore it vanishes identically on its complement in $\calss$, namely
	\begin{equation}
	\langle\sigma_1,\sigma_2\rangle=\langle\sigma_{1\,\calsm},\sigma_{2\,\calsm}\rangle\,.
	\end{equation} Finally, via the map $\lambda \mapsto \delta_\lambda \bar A_o$, which is invertible on $\calsm$, we can make the following identification:
	\begin{equation}
\boxed{	\calsm\cong T_o\calv\,.}\label{tangid}
	\end{equation}
	
	\subsubsection{The metric on the space of vacua}
	Now let us bring the metric  $\bar \rg$ in \eqref{YMmetric} in a more standard form. Remembering \eqref{lambdadef} we can write $\sigma_z=g_z^{-1}\partial_a g_z\,\dot z^a$ so that
	\begin{equation}
	\langle\sigma_z,\sigma_z\rangle=\langle e_a,e_b\rangle \dot z^a\dot z^b\,,
	\end{equation}
	where we introduced the $\calsm$-valued one-form 
	\begin{equation}
	e\equiv\left(g_z^{-1}\rd g_z\right)_\calsm\,.\label{edef}
	\end{equation}  
    The structure is now quite clear: to compute  $\bar{\rg}_z(\dot z,\dot z)$ the tangent vector $\dot z$ at an arbitrary point $z$ is first mapped to the tangent space at the reference point $T_o\calv\cong \calsm$ through $e$ and then evaluated using the fixed inner product $\langle\cdot,\cdot\rangle=(\cdot,\Dperp\cdot)$.
	
	In particular, introducing a basis\footnote{For later use it will be useful to assume the $\lambda_\ua$ to be eigenvectors of $\Dperp$, although this will not be necessary for the discussion here.} $\lambda_\ua$ for $\calsm$, we can write $e=e^\ua_a\,\lambda_\ua\,\rd z^a$ so that the metric $\bar{\rg}$ on $\calv$ takes the form
	\begin{equation}
	\boxed{\rd \bar{\mathrm{s}}^2=\bar\rg_{ab}(z)\rd z^a\rd z^b=\Dperp_{\ua\ub}\,e^\ua(z)\,e^\ub(z)\,.\label{vielbeinform}}
	\end{equation}
 	One can think of $e$ as a vielbein, since the matrix $\Dperp_{\ua\ub}\equiv(\lambda_\ua,\Dperp\lambda_\ub)$ is independent of the coordinates $z$. This particular form of the metric, \eqref{vielbeinform} together with \eqref{edef}, originates from the fact that $\calv$ is a homogeneous space as we  discuss in the following subsection.

\subsubsection{The space of vacua as a homogeneous space}
    The discussion above, where we came to a final characterization of the tangent space of $\calv$ as the subspace $\calsm\subset\calss\subset\calsg$, was based on solving the Gauss constraint and some properties of the effective Lagrangian. We saw how this was related to removing gauge transformations that act unphysically or trivially. Here we put these observations into the context of the general theory of Riemannian homogeneous spaces. Remember that our starting point was that any vacuum can be written as
    \begin{equation}
    \bar A=g_z\cdot\bar A_o\,.
    \end{equation}
    In section \ref{gaussglobal} we saw that we can restrict $g_z$ to be a representative of a class in $\cals$. There is thus a natural left action of $\cals$ on $\calv$ and because all vacua take the above form, this action is \textit{transitive}. At the same time each element of $\calv$ will be kept invariant by a subgroup $\calk_{g_z}=g_z\calk g_z^{-1}$. It is then a rather well known mathematical fact that $\calv$ is diffeomorphic to the space of right cosets 
    \begin{equation}
    \calv\cong\cals/\calk\label{quotient}\,,
    \end{equation} 
    which is also known as the \textit{orbit} $\mathcal{O}_o$ of the reference point $\bar A_o$ under the action of $\cals$. However, we know from equation \eqref{quotient_group} that $\cals$ is a quotient group  itself, hence we find that the space of physical vacua has the structure of a double quotient
    \begin{equation}
\boxed{    \calv\cong\calg_0\backslash\calg/\calk\,.\label{doublequotient}}
    \end{equation}
    It is important to point out that contrary to the first quotient by $\calg_0$, the subgroup $\calk$ is not a normal subgroup, so that $\calv$ is not a group manifold, but a \textit{homogeneous}  space (See appendix \ref{sec-homogeneous spaces} for a short review, including some of the mathematical terminology that will follow).
    
    The split \eqref{algebradecomp} of the isometry algebra $\calss$, is a standard step in the definition of a {\it reductive} homogeneous space $\cals/\calk$,  with $\calsk$ the isotropy subalgebra of a reference point, and $\calsm$ isomorphic to the tangent space at that point. The crucial property of being reductive is that 
    \begin{align}
    [\calsk,\calsm]_*\subset\calsm\,.\label{reductive property}
    \end{align}
    This is guaranteed in our case as $\calsm=\calsk^\perp$ with respect to the bi-invariant inner product \eqref{bndKilling}:
    \begin{equation}
    (\kappa_1,[\kappa_2,\lambda]_*)=([\kappa_1,\kappa_2]_*,\lambda)=(\kappa_3,\lambda)=0\,.
    \end{equation} Furthermore, as we show in appendix \ref{diffeqapp}, the symmetric operator $\Dperp$ satisfies
    \begin{equation}
    \Dperp[\kappa,\sigma]_*=[\kappa,\Dperp\sigma]_*\qquad\mbox{for all}\quad \sigma\in\calss\,,\ \kappa\in\calsk\,.\label{equivar}
    \end{equation} This is enough to guarantee that the metric \eqref{YMmetric} is invariant under the left $\cals$ action and independent of the choice of representative $g_z$, making $(\calv,\bar{\rg})$ a Riemannian homogeneous space.
    
    It is important to point out that contrary to some of the most typical examples of homogeneous spaces that appear in cosmology, supersymmetry or other areas of physics, the space of vacua $\calv$ is in general \footnote{Except in the abelian case or when $\dim M=1$.} {\it not} a symmetric space, i.e. $[\calsm,\calsm]_*\not\subset \calsk$, as we will see in the examples below. Furthermore the operator $\Dperp$ is non-trivial enough that $\calv$ is neither a naturally reductive space nor a normal reductive space. So far we have not been able to fit it inside any other well-studied subclass of reductive spaces, indeed as we will mention in section \ref{sec-geodesics} it even falls {\it outside} the class of g.o. \!spaces.

  \subsection{Representatives, coordinates and gauge fields}\label{secreps}
  In the previous subsections we discussed how the vacua of Yang-Mills theory form a homogeneous space, $\calv=\calg_0\backslash\calg/\calk$. Alternatively, and more intuitively, the vacua can be represented by choosing a particular gauge field  $\bar A(z)$ in each class of physically equivalent vacua, where $z=(z^a)$ labeling the equivalence classes, can be considered as coordinates on the space of vacua. Equivalently, the vacua $\bar A(z)$ are determined by a choice of global gauge transformation representative $g_z$ through $\bar A(z)=g_z\cdot \bar A_o$\,. 
  In this section, we will introduce such representatives, leading to well-defined, non-singular vacuum gauge fields all over spacetime. Moreover, we will present the gauge field representatives associated to curves $z(t)$ on the space of vacua, where  we will see the subtle role of the Gauss constraint.

   \subsubsection{A choice of representatives and coordinates on $\calv$}
   
   Homogeneous spaces are simple enough that the geometry all over the manifold is determined in terms of a constant inner product $\langle\cdot,\cdot\rangle$ on $\calsm\cong T_o\calv$, the tangent space at a fixed reference point $\bar A_o$. As can be seen from \eqref{vielbeinform}, the metric at other points is related to this fixed inner product through the vielbein \eqref{edef}. One can see directly from the particular form of the vielbein that it, and hence also the metric,  is invariant under the left multiplication of $g_z$ with constant group elements. Note that $e$ depends on the choice of representative $g_z$. Different choices are related as
   \begin{equation}
   g'_z=h(z) g_z k(z)\qquad \mbox{where}\quad h\in\calg_0\,,\ \ k\in\calk\,.
   \end{equation}
   Crucial to the construction is that the metric does not depend on the choice of representative, making it well defined on the quotient space. Independence of $h$ follows from the fact that the metric $\bar\rg$, see \eqref{YMmetric}, is defined completely on the boundary where $h$ goes to the identity, while invariance under $k$ follows from \eqref{equivar} and the bi-invariance of $(\cdot,\cdot)$.
   
   The practical use of a particular choice of representative depends on the situation and properties of the groups involved. Here we will  make the most straightforward choice:
   \begin{equation}
   \left.g_z\right|_{\partial M}=\left.\exp(\lambda_\ua\, z^a)\right|_{\partial M}\,.\label{boundrep}
   \end{equation}
   Here the $\lambda_\ua$ are a basis for $\calsm$, which for later convenience we will choose to be an eigenbasis of $\Dperp$, and summation is implied: $\lambda_{\ua}z^a\equiv\lambda_{\underline{1}}z^1+\lambda_{\underline{2}}z^2+\ldots$\,. For simplicity we have restricted everything to the boundary, where the exponential map is the standard one (i.e. with $G$ a matrix group, it is simply the matrix exponential).  We will discuss the extension to the bulk in the next subsection. We comment on the validity of \eqref{boundrep} in appendix \ref{diffeqapp}. 
   
   
   Using this choice of representatives, one can in principle compute the metric \eqref{YMmetric} explicitly by determining the one-form $e$ restricted to the boundary:
   \begin{eqnarray}
   \nonumber   e&\boundeq&\left(\exp(-\lambda_\ua\, z^a)\,\rd\exp(\lambda_\ua\, z^a)\right)_\calsm\\
   &\boundeq&\left(\frac{1-e^{-z^b\mathrm{ad}(\lambda_\ub)}}{z^c\,\mathrm{ad}(\lambda_\uc)}\lambda_\ua\rd z^a\right)_\calsm\,.\label{magicformula}
   \end{eqnarray}

  \subsubsection{Vacuum representatives}\label{secvacrep}
  Let us first consider the time-independent vacua $\bar A(z)$ themselves. Given \eqref{boundrep} any extension of it into the bulk can be written as\footnote{ Note that here $\exp$ is the familiar exponential map between $\calsg$ and $\calg$ which is computed by considering $\lambda\in\calsm\subset\calss\subset\calsg$, it is {\it not} the exponential map $\exp_*$ between $\calss$ and $\cals$.} $g_z=h_z \exp(\lambda_\ua z^a)$, with $h_z\in\calg_0$. Now clearly any choice of $h_z$ would provide a good representative. It seems simplest to just take $h_z=1$, so that
  \begin{equation}
  g_z=\exp(\lambda_\ua z^a)\label{bulkrepvac}\,.
  \end{equation}
  We can then finally write the gauge fields representing the vacua rather explictly:
  \begin{equation}
  \bar A(x;z)=\exp(\lambda_\ua(x) z^a)\bar A_o(x) \exp(-\lambda_\ua(x) z^a)+\exp(\lambda_\ua(x) z^a)\grad \exp(-\lambda_\ua(x) z^a)\label{explvac}\,.
  \end{equation}     
  In practice it is simplest to take $\bar A_o=0$. Remember that the $\lambda_\ua$ are defined as a basis of solutions to the linear equations $D_o^2\lambda=0$, $(\lambda,\kappa)=0$ for all $\kappa\in\ker \Dperp$, that furthermore diagonalize $\Dperp$.

  \subsubsection{Gauge field representatives for curves}\label{secadrep}
        Here we present representatives at all times for gauge fields corresponding to curves $z(t)$ on $\calv$: $A(t,x)=\bar A(x;z(t))$.  At first sight this might appear a trivial extension of the previous subsection by simply replacing $z\rightarrow z(t)$ in \eqref{explvac}, but as we'll discuss now the extension involves an additional subtlety.
        
             The Gauss constraint imposes that at all times
        \begin{equation}
        \sigma_{z(t)}=g^{-1}_{z(t)}\frac{d}{dt}g_{z(t)}\in\calss\,.
        \end{equation}
        If one would choose the representatives as in \eqref{bulkrepvac} then $\sigma_z$ would have components also along $\calsg_0$. However, we can take one step back and again consider the most general type of representative that is compatible with \eqref{boundrep}, i.e. $g_z=h_z\exp(\lambda_a)$, and choose $h_z\in \calg_0$ such that $(g^{-1}\dot g)_0=0$. This is equivalent to

        \begin{equation}
       h^{-1}_z\dot h_z=-\eta_z\,,\label{pathorederedeq}
        \end{equation}
        where
        \begin{equation}
        \eta_z=\exp(\lambda_\ua z^a)\left(\exp(-\lambda_\ua z^a)\frac{d}{dt} \exp(\lambda_\ua z^a)\right)_0\exp(-\lambda_\ua z^a)\,.\label{etaform}
        \end{equation}
        The solution to equation \eqref{pathorederedeq} is given by the path-, or in this case time-, ordered exponential 
        \begin{equation}
        h_z^{-1}=\mathrm{Pexp}\left(\int^t\eta_z dt'\right)\,.\label{hform}
        \end{equation}
        Hence we find the following representatives for adiabatic gauge fields which satisfies the Gauss constraint:
        \begin{equation}
        g_{z(t)}(x)=\left(\mathrm{Pexp}\int^t\eta_{z(t')}(x) \,dt'\right)^{-1}\exp(\lambda_\ua(x)\, z^a(t))\,.\label{timedeprep}
        \end{equation}
        It is an intriguing fact that the Gauss constraint selects a particular type of 'Wilson-line' to dress the gauge field representatives, which is path-dependent\footnote{Note that there does {\it not} exist a choice of representatives $g_z=h e^\lambda$ such that $(g_z^{-1}dg_z)_0=0$ on a patch in $\calv$ instead of just along a curve. This is because such a choice would be equivalent to $h^{-1}dh=-e^{\lambda}(e^{-\l}de^\l)_0e^{-\lambda}$, but that is impossible because the left hand side satisfies the Maurer-Cartan equations while the right hand side does so only if $[(e^{-\lambda}de^{\lambda})_\calss,(e^{-\lambda} d e^{\lambda})_\calss]_0=0$ for all $z$, which is not the case.}.
        
        Let us finally emphasize the three key features of this choice of representative. 
        \begin{enumerate}
            \item It ensures the Gauss constraint $D_o^2\sigma_z=0$.
            \item On the boundary it reduces to \eqref{boundrep} as $\eta_z$ vanishes there.
            \item For a constant curve $z(t)=z_0$ it reduces to \eqref{bulkrepvac}, as in this case $\eta_z$ is identically zero. 
            \item Surprisingly when the curve is a geodesic of the form $g(t)=\exp(t\lambda)$ (to be discussed in the following subsection) with $\lambda=\lambda_\ua v^a=const$, the dressing factor $h_z$ is again trivial. This is because for such curves $z^a(t)=v^a t$ and thus $(\exp(-\lambda t)\frac{d}{dt}\exp(\lambda t))_0=\lambda_0=0$.
        \end{enumerate}

    \subsection{Geodesics on the space of vacua}\label{sec-geodesics}
    The discussion in the previous subsection was in essence about the kinematics of our problem: how to properly restrict only to deformations along the space of vacua $\calv$. In this subsection we finally focus on the dynamics, determined by the effective Lagrangian \eqref{LeffYM}. This Lagrangian describes free motion of a point particle on the (infinite dimensional) Riemannian manifold $(\calv,\bar{\rg})$. This implies that in the adiabatic limit, solutions to the Yang-Mills equations of motion \eqref{YMeom1} reduce to geodesics on $(\calv,\bar{\rg})$.
    
    Although well-studied, the geodesic problem on a generic Riemannian manifold can be quite intractable. Here we have the advantage that $\calv=\cals/\calk$ is a homogeneous space for which the geodesic problem is more manageable, and in many cases even fully integrable. Indeed, in most well-known examples in physics the homogeneous space is ``symmetric'' or ``naturally reductive'' in which case all geodesics are simply orbits of a one-parameter subgroup of isometries: $g_{z(t)}=\exp(\sigma t)\cdot g_0$. This property is actually more general than these two subclasses and leads to the larger class of what mathematicians call {\it g.o.\,spaces} (geodesic orbit spaces). 
    There exists a simple criterion for a reductive homogeneous space to be g.o., see appendix \ref{apmath}. Interestingly $(\calv,\bar{\rg})$ falls {\it outside} this class generically. We show this explicitly for the case of non-Abelian Yang-Mills theory in 3+1 dimensions in appendix \ref{apmath}. This implies that there exist geodesics which are not generated by any isometry. Luckily there is still a large class of geodesics which do take this simple form and which we will discuss in more detail below. Finding geodesics outside this class, and determining the corresponding gauge fields, seems an interesting problem which we leave for the future.
    \subsubsection{The geodesic equation}
    Let us concretely analyze these general observations in our specific case: free motion on $\calv$, with respect to the metric \eqref{YMmetric}. The action of \eqref{LeffYM} is
    \begin{align}
    S[g(t)]=\frac{1}{2} \int dt\,\bar\rg_{ab}\dot z^a\dot z^b=\frac{1}{2} \int dt\,\langle g^{-1}\dot g,g^{-1}\dot g\rangle\label{startphys}\,,
    \end{align}
    where the coset representative $g(t)=g_{z(t)}$ is defined in \eqref{timedeprep}. The expression in the middle is the standard action for free motion of a particle on a Riemannian manifold, whose equations of motion are equivalent to the geodesic equations. The expression on the right is particular to our case where we have a homogeneous space, and it will be useful to find explicit geodesic solutions. Indeed, using this form of the action one computes that
    \begin{equation}\label{geoeq}
    \delta S=\int dt\left( -\langle g^{-1}\delta g,\dot{\sigma}_z\rangle+\langle[\sigma_z,
    g^{-1}\delta g], \sigma_z\rangle\right)\,,\qquad \sigma_z=g^{-1}\dot g\,.
    \end{equation}
    In other words, the curve $g(t)$ corresponds to a geodesic if and only if the above expression vanishes for arbitrary $\delta g$.
    
    \subsubsection{Explicit solutions}
    In the special case that we assume the geodesic to be of the form\footnote{For simplicity we restrict to geodesics starting at the reference point. For a geodesic starting at $g(0)$ one takes $g(t)=g(0)\exp(t\lambda)$.} $g(t)=\exp(t\lambda)$, with $\lambda$ constant, we have that $\sigma_z=\lambda$, and thus $\dot{\sigma}_z=0$, so that one finds the condition
    \begin{equation}
    \langle[\lambda,g^{-1}\delta g], \lambda\rangle=0\qquad\quad\mbox{for arbitrary\ }\delta g\,.\label{endphys}
    \end{equation}
    Note that, as it should, this condition coincides with the condition for a geodesic to be an isometry orbit found in the mathematical literature, see \eqref{geovec}. Using the relation \eqref{innerproducts} this is also equivalent to
    \begin{equation}
   [\lambda,\Dperp\lambda]=0\,.
    \end{equation}
    One particular set of solutions to this condition consist of the eigenvectors of the operator $\Dperp$. Choosing a basis of eigenvectors $\lambda_{\ua}\in\calsm$ one finds the following class of geodesics
    \begin{equation}
    g(t,x;v,\ua)=\exp\left(vt\,\lambda_\ua(x)\right)\label{geosol}\,.
    \end{equation}
        Note that these are not necessarily the most general geodesics of orbit form, and furthermore, as we mentioned before, there will be geodesics which are not of exponential type at all.
    
    Let us now translate the above geodesics to spacetime field configurations. Via \eqref{gauge fieldexpl} one finds
    \begin{equation}
    A_{i}(t,x;v,\ua)=g(t,x;v,\ua)\bar A_{o\,i}(x)g^{-1}(t,x;v,\ua)+g(t,x;v,\ua)\pd_ig^{-1}(t,x;v,\ua)\label{adiabaticsol}\,.
    \end{equation}
    Given the expressions for the $\lambda_\ua(x)$ the above can in principle be computed explicitly although as we will see in the examples, this might be practically complicated. It will be insightful for the following subsection to give the form of the (non-abelian) electromagnetic fields:
    \begin{equation}\label{Esol}
\boxed{    E=\dot A=-v\,g (D_o\lambda_{\ua})g^{-1}\,,\qquad B_i=\frac{1}{2}\epsilon_{ijk}F_{jk}=0\,.}
    \end{equation}
    In conclusion we see that the adiabatic solutions correspond to particular  electrostatic fields. In hindsight it is probably the reverse statement which is more interesting: certain electrostatic fields have the interpretation as a slow change of vacuum. Note the analogy with non-Abelian dyons which can be considered as slow motion on the moduli space of magnetic monopoles \cite{Julia:1975ff}. Similarly, we will show in section \ref{charges-section} that the adiabatic motions above carry various multipole charges introduced in \cite{Seraj:2016jxi}.
    
    \subsubsection{Direct route back to Yang-Mills}\label{seceoms}
    Having found the adiabatic solutions \eqref{adiabaticsol} after rephrasing the problem as a geodesic equation, it will be insightful to see how these geodesics are related to solutions of the full Yang-Mills equations of motion (\ref{Gauss},\,\ref{YMeom1}). First let us rewrite the geodesic equation \eqref{geoeq} via \eqref{innerproducts} as
    \begin{equation}
    \Dperp\dot \sigma_z+[\sigma_z,\Dperp\sigma_z]=0\,,\label{simplegeo}
    \end{equation}
    where $\sigma_z=g^{-1}_z\dot g_z$ and by construction $D_o^2\sigma_z=0$. Remembering \eqref{lambdadef} one observes that the Gauss constraint \eqref{Gauss} is exactly solved, and it also allows us to compute
    \begin{equation}
    \ddot A=-g_z(D_o\dot \sigma_z+[\sigma_z,D_o\sigma_z])g_z^{-1}\,.
    \end{equation}
    We now have all the formulae in hand to make the connection. To do so project the equations of motion \eqref{YMeom1} along the zero-mode directions, i.e. those generated by a gauge transformation such that $\delta_\gamma A=D\gamma$ and compute
    \begin{eqnarray}
\nonumber    \int_M d^3x\,\Tr\, D_i\gamma(\ddot A_i-D_jF_{ji})&=&\oint_{\partial M}\!\! d\Sigma\, n^i\, \Tr\,\gamma\left(\ddot A_i-D_{j}F_{ji}\right)-\int_M d^3 x\, \Tr\,\gamma\left(D_i\ddot A_i-D_iD_{j}F_{ji}\right)\\
\nonumber &=&\oint_{\partial M}\!\! d\Sigma\, n^i\, \Tr\,\gamma\left(\ddot A_i-D_{j}F_{ji}\right)\\
\nonumber    &=&\oint_{\partial M}\!\! d\Sigma \, \Tr\,\gamma\left(n^i\ddot A_i\right)\\
    &=&-\oint_{\partial M}\!\! d\Sigma \, \Tr\,(g_z^{-1}\gamma g_z)\left(\Dperp \dot\sigma_z+[\sigma_z,\Dperp \sigma_z]\right)\,.\label{geodesic vs eom}
    \end{eqnarray}
    To see that the volume integral in the r.h.s. of the first line vanishes, use the fact that $D_i\ddot A_i=0$ as a result of the Gauss constraint, and that $D_{i}D_{j}F_{ij}=[F_{ij},F_{ij}]=0$. To go from the second to the third line use partial integration and the boundary conditions discussed in footnote \ref{boundnote}.
    
    The computation above shows explicitely that the geodesic equation \eqref{simplegeo} implies the vanishing of those components of the Yang-Mills equations of motion \eqref{YMeom1} along the zero-mode directions and vice versa, as the above is true for arbitrary $\gamma\in\calsg$. So we conclude that the gauge fields \eqref{adiabaticsol} solve the Gauss constraint and part of the Yang-Mills equations exactly. Other components in general lead to motion in directions normal to the zero modes. However, in the adiabatic limit, they can be ignored as the effective potential becomes very steep (see footnote \ref{adiabatic limit}) and hence the geodesic motion becomes exact.
    
    We find it intriguing that according to the second line of \eqref{geodesic vs eom}, the zero mode components of the Yang-Mills field equations are nothing but the radial components of those equations on the boundary.
    
    \subsection{Multipole charges as the constants of motion}\label{charges-section}
    In this section we discuss how the conserved momenta associated with geodesics on the space of vacua and relate them to the conserved multipole charges defined in the full Yang-Mills theory. The upshot is that this way we will have presented source-free, approximate solutions of classical Yang-Mills theory carrying non-vanishing multipole charges (the nonabelian version of those defined in \cite{Seraj:2016jxi}). These solutions are interpreted as slow motion along the vacua of the theory.

    \subsubsection{Conserved momenta in the effective theory}
    For a particle freely moving on a Riemannian manifold it is well known that the momenta conjugate to isometry directions are conserved. We will compute these conserved momenta for the geodesic solutions found in the previous subsection. We then show that these momenta coincide with the multipole charges associated with the Yang-Mills solutions corresponding to these geodesics.

    The effective motion \eqref{LeffYM} is in this case on a homogeneous space $\calv\!~=~\!\cals/\calk$ equipped with the left-invariant metric $\bar{\rg}$. This implies that $\cals$ is the isometry group of the space of vacua, i.e. for each element $\sigma\in\calss$ there is a corresponding Killing vector field $\xi^a_\sigma$ of the metric $\bar{\rg}$:
    \begin{equation}
   \xi^a_\sigma=(g_z^{-1}\sigma g_z)^\ua\,e_\ua^a\,, \label{Killingv}
    \end{equation}
    where
    \begin{equation}
    (g_z^{-1}\sigma g_z)_\calsm=(g_z^{-1}\sigma g_z)^\ua\lambda_\ua\,,
    \end{equation}
    and $e_\ua^a$ is the inverse of the vielbein \eqref{edef}, i.e. $ e_{\ua}^a\,e_a^\ub=\delta_\ua^\ub$. For a nice review on the construction of the above Killing vectors see e.g. \cite{Castellani:1983tb}.
    
    One can then construct the following momenta $P_\sigma$ conjugate to the Killing vectors \eqref{Killingv} which are constants of geodesic motion:
    
    \begin{align}
    P_\sigma&=\dfrac{\pd L}{\pd \dot{z}^a}\xi^a[\sigma]=\bar{\rg}_{ab}\dot{z}^a\xi^b[\sigma]=(g_z^{-1}\sigma g_z,\Dperp \sigma_z)\,.
    \end{align}
    Evaluating these momenta for the geodesic solutions \eqref{geosol} passing through the reference point, one finds that the charges vanish for $\sigma\in\calsk$, while when $\sigma\in\calsm$ one has the following independent set of non-vanishing charges
    \begin{equation}
    P_{\lambda_{\ua}}[g(v,\ub)]=v\,\Dperp_{\ua\ub}\,.
    \end{equation}
 where $\Dperp_{\ua\ub}=(\lambda_\ua,\Dperp\lambda_\ub)$ are the components of the symmetric operator $\Dperp$ on $\calsm$.
    
    \subsubsection{Conserved charges in the full theory}
    The conserved charges associated with global gauge symmetries of Yang-Mills theory can be computed using the covariant phase space method. This is discussed in appendix \ref{app-charges}. Writing out the definition \eqref{charges-YM} in our particular case gives
    \begin{equation}
    Q_{\gamma}[A]=-\oint_{\partial M}\!\!d\Sigma\, n^i\,\Tr\,E_i\gamma\,.
    \end{equation}
    These are a non-abelian generalization of the electric multipole charges of \cite{Seraj:2016jxi}. It is a gratifying that the conserved momenta $P_\sigma$ of the effective theory coincide with the conserved charges $Q_\sigma$ of the full Yang-Mills theory:
    \begin{eqnarray}
\nonumber    P_\sigma&=&(g_z^{-1}\sigma g_z,\Dperp\sigma_z)\\
\nonumber    &=&\oint_{\partial M}d\Sigma\, \Tr\,g_z^{-1}\sigma g_z\Dperp\sigma_z\\
\nonumber    &=&\oint_{\partial M}d\Sigma\, n^i\,\Tr\,\sigma g_z(D_{o\,i}\sigma_z)g_z^{-1}\\
&=&-\oint_{\partial M}d\Sigma\, n^i\,\Tr\,E_i\sigma\,,
    \end{eqnarray}
    where in the last step we used the expression \eqref{lambdadef} for $E=\dot A$, valid for gauge fields corresponding to motion on the space of vacua $\calv$.
    Therefore we conclude that 
    \begin{align}
    \boxed{P_\sigma=Q_\sigma \,,\qquad \forall \sigma\in \calss\,.}
    \end{align} 
   
    The Poisson bracket of two charges can also be computed using \eqref{Poisson bracket}
\begin{align}
\nonumber    \{Q_{\gamma_1},Q_{\gamma_2}\}&=\de_{\gamma_2}Q_{\gamma_1}=-\oint_{\pd\Sigma} d\Sigma_\mn\, \Tr\, [F^\mn,\gamma_2] \gamma_1\\
    &=-\oint_{\pd\Sigma} d\Sigma_\mn\, \Tr\, F^\mn [\gamma_2,\gamma_1]=Q_{[\gamma_1,\gamma_2]}\,.
\end{align}
Note that the charges are defined for any gauge parameter $\gamma\in\calsg$, but that they vanish for all local gauge transformations, i.e. $Q_{\kappa}=0$ when $\kappa\in \calsg_0$. As the charges are linear in the gauge parameters it follows that $Q_{\gamma}=Q_{\gamma+\calsg}=Q_{\gamma_\calss}$, where we remind that $\gamma_\calss$ is the component of $\gamma$ along $\calss$, i.e. the unique element in  $\calss$ such that $\left.\gamma_\calss\right|_{\partial M}=\left.\gamma\right|_{\partial M}$. This implies that any type of non-trivial conserved multipole charge can be carried by our adiabatic solutions, i.e. $Q_\gamma=P_{\gamma_\calss}$.
 
Finally let us point out that the algebra of charges is isomorphic exactly to the algebra of global gauge symmetries defined through the modified bracket \eqref{YMcom}:
\begin{align}
    \{Q_{\sigma_1},Q_{\sigma_2}\}=Q_{[\sigma_1,\, \sigma_2]_*}\,.
\end{align}    
Note that this algebra can also be interpreted as the algebra of boundary gauge transformations, similar to the infinite volume case \cite{Strominger:2013lka, Barnich:2013sxa}.

    \section{Examples}\label{exampsec}
    In this section we illustrate the general discussion of the previous section in a few interesting exemplary cases.
    
    First we discuss Maxwell theory, where due to the Abelian nature of the gauge group many things simplify. This allows us to work out all objects very concretely. Although the geometry of the space of vacua in that example turns out to be flat, the main aim  is to illustrate the physical interpretation of the geodesic solutions: they are source-free electrostatic solutions.
    
     In the second example we discuss Yang-Mills theory with a non-Abelian gauge group in 3+1 dimensions, which is the example of most direct physical interest. In that case, however, things are more complicated and we will not be able to give for example the metric in a fully explicit form. For this reason we will also give a third example, namely non-Abelian gauge theory in 1+1 dimensions. This example has the advantage of being richer than Maxwell theory, with a non-trivial, curved space of vacua, while being fully computable, which makes it quite illustrative.
    
    Finally let us point out that in the previous section we kept the reference vacuum $\bar A_o$ arbitrary, to emphasize that there is a priori no preferred vacuum. In practice it is of course simplest to take $A_o=0$, which we will do in all examples below, so that $D_0=\grad$.
    
        \subsection{Maxwell theory}
            In our first example we consider Maxwell's electrodynamics, which fits in our general discussion by taking the gauge group $G=$U(1). We will restrict attention to $3+1$ dimension with the spatial manifold a flat and finite ball of radius $R$, i.e. $M=B_R^3\subset\mathbb{R}^3$ and $\partial M=S^2_R$. Much of our discussion will be in spherical coordinates, with $r$ the radial direction. This example is simple enough that we can write out all objects concretely and we will take the time to spell out a number of the constructions above in full detail which might help in clarifying the discussions of section \ref{mainsec}.
            
            \subsubsection{Gauge algebra decomposition}
            Let us start by discussing the split of the Lie algebra of infinitesimal gauge transformations $\calsg=\calsg_0\oplus\calss$ and $\calss=\calsk\oplus\calsm$. The gauge parameters $\gamma$ are in this case purely imaginary functions\footnote{Note that we are using so called geometric conventions, where the gauge field and gauge parameters are anti-Hermitian, instead of the convention to choose them Hermitian that is often used in physics. Related to this, remember that $\Tr=-\tr$ so that in case of Maxwell theory it reduces to a factor of $-1$.} on $B_R^3$, i.e. $\gamma(r,\theta,\phi)\in i\mathbb{R}$.
          	Choosing $\bar A_o=0$ the Gauss constraint \eqref{YMGauss} reduces to $\partial_i\partial_i \sigma=0$ whose non-singular solutions form the subspace $\calss$, which thus has the following basis 
          	\begin{align}
            \sigma_{\ell m}\equiv i\left(\frac{r}{R}\right)^\ell Y_{\ell m}\,,
          	\end{align}
          	where $Y_{\ell m}(\theta,\phi)$ are {\it real} spherical harmonics. 
          	Let us use this explicit construction of $\calss$ to illustrate the split $\gamma=\gamma_0+\gamma_\calss$. Remember that $\gamma_\calss$ is defined as the unique element in $\calss$ such that $\left.\gamma\right|_{\partial M}=\left.\gamma_\calss\right|_{\partial M}$, whose unique solution is 
          	\begin{equation}
          	\gamma_\calss(x)=\frac{1}{R^2}\sum_{\ell m}(\gamma,\sigma_{\ell m})\,\sigma_{\ell m}(x)\qquad\mbox{with}\quad (\gamma,\sigma_{\ell m})=-iR^2\oint_{S^2_R}\! d\Omega\, \gamma\, Y_{\ell m}\,.\label{maxproj}
          	\end{equation}
          	Then of course $\gamma_0=\gamma-\gamma_\calss$, which is easily seen to vanish on the boundary and hence an element of $\calsg_0$. Remembering that the group of global gauge transformations $\cals$ is isomorphic to the group of boundary gauge transformations $\left.\calg\right|_{S^2}$ we see that here $\cals$ is isomorphic to the abelian group of all maps $S^2\rightarrow {\mathrm{U}}(1)=S^1$ , with $\calss$ simply being the abelian algebra of imaginary functions on the sphere. In this case the modified bracket $[\cdot,\cdot]_*$ is trivially equal to the commutator, as they both identically vanish.
          	
          	Let us continue by working out the operator $\Dperp$, defined in \eqref{Dperpdef}.
            Using \eqref{maxproj}, we find its action to be
          	\begin{equation}
          	\Dperp\sigma=\frac{1}{R}\sum_{\ell m}\,\ell\,(\sigma,Y_{\ell m})\,\sigma_{\ell m}\label{DperpMaxwell}\,.
          	\end{equation}
          	 It is an intriguing observation that this operator becomes the dilatation operator:
          	 \begin{equation}
          	 \Dperp\sigma=\frac{r}{R}\partial_r \sigma\,.
          	 \end{equation}

          	The kernel of $\Dperp$ defines the subspace $\calsk$. From \eqref{DperpMaxwell} we see that this is the $\ell=0$ subspace of $\calss$, generated by $\sigma_{0\,0}=1/(2\sqrt{\pi})$. Equivalently, elements $\kappa\in\calsk$ can be represented by the condition $\partial_i\kappa=0$. This then illustrates the observation we made above that $\calsk$ is also the algebra of isotropic gauge transformations, since in Maxwell theory $\delta_\gamma A_i=\partial_i\gamma$, and hence the constant gauge transformations act trivially.
          	
          	Now remember that $\calsm=\calsk^\perp$, so we see that we can identify its basis $\lambda_{\ua}$, via $\ua=(\ulm)$ for $\ell\geq 1$, with          	     	
          	\begin{align}
            \lambda_{\ulm}\equiv\left(\frac{r}{R}\right)^\ell Y_{\ell m}\,,\qquad \ell\geq 1\,.
          	\end{align}
          	Note that these are also eigenvectors of the operator $\Dperp$:
          	\begin{equation}
          	\Dperp \lambda_{\ulm}=\frac{\ell}{R}\lambda_{\ulm}\,,
          	\end{equation}
          	so that
          	\begin{eqnarray}
          	\Dperp_{\ulm\,\ulmp}\equiv(\lambda_{\ulm},\Dperp\lambda_{\ulmp})=R\,\ell\,\delta_{\ell\,\ell'}\delta_{m\,m'}\,.\label{MDmat}
          	\end{eqnarray}
          	
          	\subsubsection{Geometry}
	        The formula \eqref{MDmat} immediately leads to the metric on the space of vacua $\calv$ in the form \eqref{vielbeinform}:
	        \begin{equation}
	        \bar{\rg}_{\ell m\,\ell'm'}(z)=R\sum_{\ell''m''}\ell''\,e_{\ell m}^{\underline{\ell''m''}}\,e_{\ell' m'}^{\underline{\ell''m''}}\,.\label{Mmetric1}
	        \end{equation}
	        The remaining step is then to compute the vielbein $e$ on the boundary. We remind the definition \eqref{edef} and our particular choice of representatives \eqref{boundrep}, that define the coordinates $z^a=(z^{\ell m})$. It then follows that on $\pd M=S^2_R$
	        \begin{equation}
	        g_z\boundeq\exp\Big(i\sum_{\ell\geq 1, m}Y_{\ell m}z^{\ell m}\Big)\,,
	        \end{equation}
	        and thus
	        \begin{equation}
	        e\boundeq i \sum_{\ell\geq 1, m}Y_{\ell m}\,\rd z^{\ell m}\,.
	        \end{equation}
	        Note that in this special case $e$ is constant on the moduli space, i.e. $z$-independent. Consequently the metric \eqref{Mmetric1} is flat and in these coordinates takes the simple form
	        \begin{equation}
	        \rd\bar{\mathrm{s}}^2=\bar{\rg}_{\ell m\,\ell'm'}\,\rd z^{\ell m}\rd z^{\ell'm'}=R\sum_{\ell\geq 1, m}\!\ell\, \rd z^{\ell m}\rd z^{\ell m}\,. \label{Mmetric2}
	        \end{equation}
	        \subsubsection{Adiabatic solutions}          
          	The effective Lagrangian \eqref{LeffYM} with the metric \eqref{Mmetric2} is rather trivial, describing an infinite number of non-interacting free particles:
          	\begin{equation}
          	L=\frac{1}{2}R\sum_{\ell\geq 1, m}\!\ell\, (\dot z^{\ell m})^2\,.
          	\end{equation}
          	The corresponding motion or geodesics are just the straight lines 
          	\begin{equation}
          	z^{\ell m}(t)=v^{\ell m}t+z_0^{\ell m}\,.
          	\end{equation}
           It follows that for this motion the representatives \eqref{timedeprep} reduce to
           \begin{equation}
           g_{z(t)}=\exp\Big(i\sum_{\ell\geq 1, m}\left(\frac{r}{R}\right)^\ell Y_{\ell m}\,(v^{\ell m}t+z_0^{\ell m})\Big)\,,
           \end{equation}
           so that
           \begin{equation}
           A(t,x)=\bar A(x;z(t))=g_{z}dg_z^{-1}=-i\sum_{\ell\geq 1, m}(v^{\ell m}t+z_0^{\ell m})\,d\left(\left(\frac{r}{R}\right)^\ell Y_{\ell m}\right)\,.
           \end{equation}
           This gauge field leads to a vanishing magnetic field $B_i=0$ and an electric field\footnote{Here for clarity we go back to physics conventions choosing the electric field to be purely real, rather than purely imaginary, $E_{\mathrm{phys}}=-i E_\mathrm{geom}=-i\dot A$.}
          	\begin{align}
          	E_i=-\pd_i\Phi, \qquad \Phi=\sum_{\ell\geq 1, m}v^{\ell m}\left(\frac{r}{R}\right)^\ell Y_{lm}\label{Efields}  \,.
          	\end{align}
          	
          	These are nothing but the source free electrostatic solutions of Maxwell theory inside the ball. We find it a very interesting observation that these electrostatic solutions can be interpreted as geodesic motion on the space of vacua of the theory. This provides an additional, very physical, argument that indeed global gauge transformations should be considered as generating physically inequivalent configurations and a non-trivial space of vacua.
          	
          	Note that in this Abelian example the adiabatic approximation turns out to be exact and the solutions obtained remain uncorrected even for high velocities. This is because Maxwell theory is a linear theory and the zero-modes are decoupled from the non-zero-modes, so that it is consistent to put the latter to zero.
          	
          	Finally let us comment on another reason why these electrostatic fields are of interest. In \cite{Seraj:2016jxi} it was shown that the well-known multipole moments associated to a charged matter distribution can be completed into multipole charges that are conserved by associating a particular multipole charge to the electromagnetic field itself. This construction implies there should also exist matter free, pure field configurations that carry these charges. Indeed, the solutions \eqref{Efields} provide such pure field configurations. As we discussed in section \ref{charges-section}, the charges associated to the adiabatic solutions can equivalently be interpreted as the conserved momenta of the effective particle motion, which in this case take the simple form
          	\begin{equation}
          		Q_{\lambda_\ulm}=P_{\ell m}=R\,\ell\, \dot z^{\ell m}=R\,\ell\, v^{\ell m}  \,.
          	\end{equation}
          	Note that the total electric charge corresponding to $\ell=0$,  vanishes as it should for a pure field configuration. Here we see this is directly related to the isotropic gauge transformations that have been quotiented out.

    \subsection{Non-Abelian Yang-Mills}
    Now we move beyond Maxwell theory by considering the gauge group $G$ to be a non-abelian (compact, semi-simple) Lie group. As before we assume also in this example $M=B_R^3\subset\mathbb{R}^3$ and $\partial M=S_R^2$. Many things will be almost similar to the Maxwell discussion, except for the appearance of the gauge group generators $T_I$, which form a basis for $\mathfrak{g}$. The fact that these do not commute, however, leads to a much richer structure. 
    
    \subsubsection{Gauge algebra decomposition}\label{4dYMsplit}
    Because of the choice of reference vacuum $\bar A_o=0$, the operator $D_o^2=\partial_i\partial_i$ remains like before the 3d Laplacian, so that we have the following basis for $\calss$:
    \begin{equation}
    \sigma_{I\,\ell m}\equiv T_I\left(\frac{r}{R}\right)^{\ell} Y_{\ell m}\label{sbasis}  \,.
    \end{equation}
    The projection \eqref{maxproj} generalizes to
    \begin{equation}
    \gamma_\calss=\frac{1}{R^2}\sum_{I\,\ell m}(\gamma,T_I\,Y_{\ell m})\,\sigma_{I\,\ell m}\qquad\mbox{with}\quad (\gamma,T_I\,Y_{\ell m})=R^2\oint_{S^2_R}\! d\Omega\,\Tr\, \gamma\, T_I\,Y_{\ell m}\,.\label{YMproj}
    \end{equation}
    Now the group of global gauge transformations $\cals$ is non-abelian and hence the bulk representation of the algebra $\calss$ has a bracket which is a non-trivial modification of the commutator. Using the definition \eqref{YMcom} and the formula \eqref{YMproj} for the projector, one computes
    \begin{equation}
    [\sigma_{I\,\ell m},\sigma_{J\,\ell'm'}]_*=\sum_{K\,\ell''m''}f_{IJ}{}^{K}\,C_{\ell m\,\ell'm'\,\ell''m''}\,\sigma_{K\,\ell''m''}\label{salgebra}  \,,
    \end{equation}
    where\footnote{Note that the real Gaunt coefficients $C_{\ell m\,\ell'm'\ell''m''}$ satisfy slightly different selection rules than their complex analogues which are more often used. See \cite{Stein} for a detailed analysis of the real Gaunt coefficients and the relation to the complex case.}
    \begin{equation}
    [T_I,T_J]=\sum_K f_{IJ}{}^{K}T_K\qquad\quad C_{\ell m\,\ell'm'\ell''m''}=\oint_{S^2}\!\!d\Omega\, Y_{\ell m}Y_{\ell' m'}Y_{\ell''m''}  \,.
    \end{equation}
    Note that, in accordance with the general discussion of  section \eqref{sec-ASG}, the algebra \eqref{salgebra} is isomorphic to that of the boundary gauge transformations.
    
    The operator $\Dperp$ is essentially identical to the Maxwell example, with
    	\begin{equation}
    	\Dperp\sigma=\frac{1}{R}\sum_{I\,\ell m}\,\ell\,(\sigma,T_I\,Y_{\ell m})\,\sigma_{I\, \ell m}\qquad\Leftrightarrow\quad \Dperp=\frac{r}{R}\partial_r\label{DperpYM}\,.
    	\end{equation}
    Note that also in this case $\calsk=\ker\Dperp=\mathfrak{g}$ is still given by the constant gauge transformations and a natural basis for its orthogonal complement $\calsm$ that diagonalizes $\Dperp$ is given by
    \begin{equation}
    \lambda_{\uI\,\ulm}=T_I \left(\frac{r}{R}\right)^{\ell} Y_{\ell m}\,,\quad\ell\geq 1\qquad\mbox{with}\quad \Dperp \lambda_{\uI\,\ulm}=\frac{\ell}{R} \lambda_{\uI\,\ulm}\,.\label{eigeq}
    \end{equation}
    It follows that 
    \begin{eqnarray}
    \Dperp_{\uI\,\ulm\ \uJ\,\ulmp}=R\,\mathbb{G}_{IJ}\,\ell\,\delta_{\ell\ell'}\delta_{mm'}\label{YMDmat}  \,,
    \end{eqnarray}
    with $\mathbb{G}_{IJ}=\Tr\,T_IT_J$ the Killing form on $\mathfrak{g}$.
    
    \subsubsection{Geometry}
    The line-element in vielbien form \eqref{vielbeinform} following from \eqref{YMDmat} becomes
    \begin{equation}
    \rd\bar{\mathrm{s}}^2=R\!\!\sum_{I,J,\ell\geq 1, m}\mathbb{G}_{IJ}\,\ell\,e^{\uI\,\ulm}\,e^{\uJ\,\ulm}  \,.
    \end{equation}
    Here is where the first crucial difference with the Maxwell example appears. Contrary to that example, where the constant vielbeins $e^\ua$ led to a flat metric, here the vielbein $e$ is highly non-trivial. Actually we have not been able to work it out beyond its definition:
    \begin{equation}
    e=\left(g_z^{-1}\rd g_z\right)_\calsm\,,\qquad\qquad \left.g_z\right|_{S_R^2}=\exp\left(\sum_{I,\ell\geq1,m}T_{I}Y_{\ell m}\,z^{I\,\ell m}\right)  \,.
    \end{equation}
    In principle one could try to compute this using the formula \eqref{magicformula}, but successive powers of the adjoint representation, essentially given in \eqref{salgebra}, seem to become intractable. Maybe this problem can be solved using a different approach or another choice of coordinates, which could be interesting to attempt in the future. The curvature and other geometric invariants of homogeneous spaces can be computed purely algebraically, see e.g. \cite{arvanitogeorgos2003introduction}, it might be interesting to attempt this for the geometry above.
    
    \subsubsection{Adiabatic solutions}
    As we show in appendix \ref{aphom} in the example considered here the homogeneous space $(\calv,\bar\rg)$ is {\it not} g.o., so that there exist geodesics which are not orbits. However, as we showed in section \ref{sec-geodesics}, there is still a large number of geodesics which are orbits, see \eqref{geosol}, generated by eigenvectors of $\Dperp$. From \eqref{eigeq} it follows that the linear combination $\sum_{I,m}\,v^{I\,\ell m} \lambda_{\uI\,\ulm}$ is an eigenvector for any fixed $\ell$. For this choice the corresponding electric field satisfies
    \begin{equation}
     E_i=- g_z \,\partial_i\Phi \,g_z^{-1}, \qquad \Phi=\sum_{I,m}\,v^{I\,\ell m}\left(\frac{r}{R}\right)^\ell Y_{\ell m}\,T_I\label{EfieldsYM}  \,.
    \end{equation}
    This is an expression analogous to the Maxwell case \eqref{Efields} and the physical interpretation is identical to that discussed there . Also the analysis of the conserved charges is parallel to the Maxwell example and follows the general discussion of section \ref{charges-section}, so we will end our discussion here.

    \subsection{A toy example in 1+1 dimensions}\label{toyex}
    In 1+1 dimensions the spatial manifold is one-dimensional and the analogue of a ball is an interval, i.e. $M=(-R,R)$, which adds the small twist that the boundary is disconnected and made of two points, $\partial M=\{-R,R\}$. This example is of interest because a zero-dimensional boundary will lead to a finite dimensional space of vacua and it renders this example fully tractable even when the gauge group $G$ is non-Abelian. At the start  $G$ is arbitrary, but later we will restrict to the case $G=$SU(2) for simplicity.
    
    \subsubsection{Gauge algebra decomposition}
    As in other examples, we make the choice $\bar A_o=0$, so that $\calss$ is the space of non-singular solutions to $\frac{d^2}{dx^2}\sigma=0$. The unique element $\sigma\in\calss$ with the boundary values $\sigma(-R)=\sigma_-$ and $\sigma(R)=\sigma_+$ is then
    \begin{equation}
    \sigma=\frac{\sigma_+-\sigma_-}{2}\frac{x}{R}+\frac{\sigma_++\sigma_-}{2}  \,.
    \end{equation}
    Note that  $\sigma_\pm\in \mathfrak{g}$ so that $\calss\cong\mathfrak{g}\times\mathfrak{g}=\left.\calsg\right|_{\partial M}$.  As this second way of representing things is simpler and more intuitive we will sometimes write $\sigma\cong(\sigma_-,\sigma_+)$. 
    From the characterization above, it follows that for an arbitrary gauge transformation $\gamma(x)\in\calsg$, the projection on global gauge transformations is simply
    \begin{equation}
    \gamma_\calss(x)=\frac{\gamma_+-\gamma_-}{2}\frac{x}{R}+\frac{\gamma_++\gamma_-}{2}  \,.    \label{1dproj}\end{equation}
    One can then compute the modified bracket \eqref{YMcom}
    \begin{equation*}
    [\sigma_1,\sigma_2]_*=\frac{[\sigma_{1+},\sigma_{2+}]-[\sigma_{1-},\sigma_{2-}]}{2}\frac{x}{R}+\frac{[\sigma_{1+},\sigma_{2+}]+[\sigma_{1-},\sigma_{2-}]}{2}\cong ([\sigma_{1-},\sigma_{2-}],[\sigma_{1+},\sigma_{2+}])  \,.
    \end{equation*}
   Again we see that the modified bracket provides an extension to the bulk of the algebra of boundary gauge transformations. 
   
   Next we compute the operator $\Dperp$ defined in \eqref{Dperpdef}:
    \begin{equation}
    \Dperp\sigma=\frac{\sigma_+-\sigma_-}{2R}\frac{x}{R}\cong \left(\frac{\sigma_--\sigma_+}{2R},\frac{\sigma_+-\sigma_-}{2R}\right)  \,.
    \end{equation}
   Note that  also in this example $\Dperp$ reduces to the dilatation operator:
    \begin{equation}
    \Dperp\sigma=\frac{x}{R}\frac{d}{dx}\sigma  \,.
    \end{equation}
    Via the definition $\calsk=\ker \Dperp$  we can identify it as the constant gauge transformations which indeed coincide with the isotropic gauge transformations $\delta_\kappa\bar A_o=0$. Furthermore, as for constant gauge transformations $\kappa_+=\kappa_-=\kappa$ we have the identification
    \begin{equation}
    \calsk=\{\kappa\in\calss\,|\, \frac{d}{dx}\kappa=0\}\cong\{(\kappa,\kappa)\,|\, \kappa\in \mathfrak{g}\}=\mathfrak{g}_\mathrm{diag}  \,.
    \end{equation}
    So in this simple example the space of vacua $\calv$ becomes a rather standard homogeneous space
    \begin{equation}
    \calv=\cals/\calk\cong(\mathfrak{g}\times\mathfrak{g})/\mathfrak{g}_\mathrm{diag}  \,.
    \end{equation}
    The inner product on $\calss$ given in \eqref{bndKilling} reduces in this example to
    \begin{equation}
    (\sigma_1,\sigma_2)=\frac{1}{2}\Tr (\sigma_{1-}\sigma_{2-}+\sigma_{1+}\sigma_{2+})    \,,
    \end{equation}
    so that
    \begin{equation}
    \langle \gamma_1,\gamma_2\rangle=(\gamma_1,\Dperp \gamma_2)=\frac{1}{4R}\Tr (\gamma_{1+}-\gamma_{1-})(\gamma_{2+}-\gamma_{2-})\,.
    \end{equation}
    
    Remembering that the subspace $\calsm$ is then defined as the orthogonal complement of $\calsk$ with respect to $(\cdot,\cdot)$ one finds that
    \begin{equation}
    \calsm=\{ \lambda\,|\, \lambda=\lambda_+\frac{x}{R}\,,\ \lambda_+\in\mathfrak{g}\}\cong \{(-\lambda_+,\lambda_+)\,|\,\lambda_+\in\mathfrak{g}\}\,.
    \end{equation}
    We here see clearly that $\langle\cdot,\cdot\rangle$, which in general is degenerate, becomes positive definite when restricted to $\calsm$:
    \begin{equation}
    \langle\lambda,\lambda\rangle=\frac{1}{R}\Tr\,\lambda_+^2\geq 0\qquad \forall\lambda\in\calsm\,.
    \end{equation}
    Observe that the split $\calss=\calsk\oplus\calsm$ is indeed reductive, since
    \begin{equation}
    [\calsk,\calsm]_*\subset\calsm\qquad\mbox{via}\quad [(\kappa,\kappa),(-\lambda_+,\lambda_+)]_*=(-[\kappa,\lambda_+],[\kappa,\lambda_+])\,.
    \end{equation}
    Interestingly, contrary to the higher dimensional case, in this 1+1 dimensional example the space of vacua $\calv$ will actually be a symmetric space because
    \begin{equation}
    [\calsm,\calsm]_*\subset\calsk\qquad\mbox{via}\quad [(-\lambda_{1+},\lambda_{1+}),(-\lambda_{1+},\lambda_{1+})]_*=([\lambda_{1+},\lambda_{2+}],[\lambda_{1+},\lambda_{2+}])\,.
    \end{equation}
    
    \subsubsection{Local geometry}
    As a vector space $\calsm$ is isomorphic to $\mathfrak{g}$ (through $\calsm$ is not an algebra) and so a basis $T_I$ of $\mathfrak{g}$ also provides a basis $\lambda_\ua$ of $\calsm$:
    \begin{equation}
    \lambda_\uI=T_I\frac{x}{R}\,.
    \end{equation}
    The choice of coset representative \eqref{boundrep} thus becomes
    \begin{equation}
    \left.g_z\right|_{\partial M}\cong (g_z(-R),g_z(R))=(\exp(-T_I z^I),\exp(T_I z^I))  \,.
    \end{equation}
    Note that as we have a real coordinate $z^I$ for each generator $T_I\in\mathfrak{g}$ as a manifold $\calv$ will be {\it locally} diffeomorphic to $G$, but as we will see below it will differ globally and geometrically.
    To compute the vielbein $e$ we first compute the left-invariant form $\theta=g^{-1}_z\rd g_z$:
    \begin{equation}
    \left.\theta\right|_{\partial M}=(\theta_-,\theta_+)\,,
    \end{equation}
    where $\theta_+$ is the Maurer-Cartan form on $\mathfrak{g}$:
    \begin{equation}
    \theta_+=\exp(-T_I z^I)\rd \exp(T_I z^I)\quad\mbox{and}\quad \theta_-=-\exp(T_I z^I)\,\theta_+ \exp(-T_I z^I)\,.
    \end{equation}
    Now note that $\theta$ has components both along $\calsk$ and $\calsm$, with $\theta_\calsm=e$\,:
    \begin{equation}
    \theta_\calss=\theta_{\calsk}+\theta_{\calsm} \cong\left(\frac{\theta_++\theta_-}{2},\frac{\theta_++\theta_-}{2}\right)+\left(\frac{\theta_--\theta_+}{2},\frac{\theta_+-\theta_-}{2}\right)\,.
    \end{equation}
    To continue    It then follows that the line element on $\calv$ is
    \begin{equation}
    \rd\bar{\mathrm{s}}^2=\langle e,e\rangle=\frac{1}{2R}\left(\Tr\,\theta_+^2-\Tr\,\theta_+\theta_-\right)\label{1dmet}\,.
    \end{equation}
    Note that the first term is the standard bi-invariant metric on the Lie group $G$.
        
    To continue we choose the particular case $G=$SU(2). In this case a basis for the Lie algebra $\mathfrak{su}(2)$ is given by the generators $T_I=\frac{i}{2}{\bm \sigma}_I$, so that we can parameterize a general element $\lambda_+\in\mathfrak{su}(2)$ as
    \begin{equation}
    \lambda_+=2\psi\, n^IT_I\,,\qquad n=(\sin\vartheta\cos\varphi,\sin\vartheta\sin\varphi,\cos\vartheta)\,.
    \end{equation} 
    These are related to the coordinates used throughout the paper as $z^I=2\psi\, n^I$. This choice of coordinates leads to the standard parameterization of SU(2):
    \begin{equation}
    \exp(\lambda_+)=\cos\psi\,\mathbf{1}+i\sin\psi\,n^I{\bm \sigma}_I\label{su2}\,.
    \end{equation}
    In turn one finds by direct computation that
    \begin{eqnarray}
    \theta_+&=&i{\bm \sigma}_I\left(n^I\rd\psi+\sin\psi\,(\cos\psi\,\rd n^I+\sin\psi\,\epsilon_{IJK}\,n^J\,\rd n^K)\right)\\
    \theta_-&=&-i{\bm \sigma}_I\left(n^I\rd\psi+\sin\psi\,(\cos\psi\,
    \rd n^I-\sin\psi\,\epsilon_{IJK}\,n^J\,dn^K)\right)\,,
    \end{eqnarray}
    so that the metric \eqref{1dmet} is
    \begin{eqnarray}
    2R\,\rd\bar{\mathrm{s}}^2=&=&4\rd\psi^2+\sin^22\psi\,(\rd\vartheta^2+\sin^2\vartheta\,\rd\varphi^2)\\
    &=&d\chi^2+\sin^2\chi\,(d\vartheta^2+\sin^2\vartheta\,d\phi^2)\label{roundS3}\,.
    \end{eqnarray}
    In the second line, after the redefinition $\chi=2\psi$ we recognize the round metric on S$^3$. At first this might appear surprising, as the three-sphere with the round metric corresponds to the group manifold SU(2) with its bi-invariant metric, but the space of vacua $\calv$ is not naturally a group manifold. It will be interesting to explain this in some more detail, as this will also illustrate the global structure of $\calv$, something we did not analyze in the other examples. 

\subsubsection{Global geometry}    
    The global structure of $\calv$ is encoded in the precise ranges of the coordinates $\chi, \vartheta$ and $\varphi$. To understand those one needs to go back to the expression \eqref{su2}.
    An arbitrary special unitary matrix can be reached by choosing the following ranges for the angles: $\varphi\in[0,2\pi)$, $\vartheta\in[0,\pi]$ and $\psi\in[0,\pi]$. But we should be careful, our aim is {\it not} to parameterize an arbitrary element of SU(2), rather we would like to parameterize all cosets that make up $\calv=(\mathrm{SU}(2)\times\mathrm{SU}(2))/\mathrm{SU}(2)_\mathrm{diag}$. To do so we used the coset representatives $(\exp(-\lambda_+),\exp\lambda_+)$. At first this might appear the same as parameterizing elements $\exp(\lambda_+)$ of SU(2), but the subtle difference is that this choice of representatives is not globally well defined. Indeed the two seemingly 'different' representatives $(\exp(-\lambda_+),\exp(\lambda_+))$ and $(-\exp(-\lambda_+),-\exp(\lambda_+))$ are actually elements of the same coset! So the pairs of the form $(\exp(-\lambda_+),\exp(\lambda_+))$ do not form a unique set of representatives on all of the homogeneous space (while they are good representatives close to the identity). In the case at hand one can verify that this problem of having multiple representatives for the same coset is restricted to the degeneracy of $(\exp(-\lambda_+),\exp(\lambda_+))$ and $(-\exp(-\lambda_+),-\exp(\lambda_+))$ only. Furthermore, using \eqref{su2}, one sees that under $\psi\rightarrow \pi-\psi, \vartheta\rightarrow \pi-\vartheta, \varphi\rightarrow \pi+\varphi$ we have that $\exp(\lambda_+)\rightarrow -\exp(\lambda_+)$. So we can make a unique choice of representatives by restricting $\psi\in [0,\frac{\pi}{2}]$. Note that generically this degeneracy is two-fold, except when $\psi=\frac{\pi}{2}$ as in this case for all $\vartheta, \varphi$ we have $\exp(\lambda_+)=in^I{\bm \sigma_I}$ and thus $\exp(-\lambda_+)=-\exp(\lambda_+)$. Note this nicely fits with the geometry \eqref{roundS3}, where $\psi=\frac{\pi}{2}$, i.e. $\chi=\pi$ corresponds to the south pole which is indeed a single point for all $\vartheta, \varphi$. The conclusion of this discussion is that we should take $\chi\in[0,\pi]$ so that indeed $\calv\cong\, $S$^3$. Pictorially $\calv$ is obtained by cutting the original S${}^3$=SU(2) along the equator, keeping one half, identifying the full equator as one point\footnote{Note here the crucial difference with the construction of SO(3), which can be obtained by cutting S${}^3$ into half but then closing the equator by identifying opposite points.} and then putting on that space, which again topologically is a 3-sphere, the round metric so that
    \begin{equation}
    (\calv,\bar\rg)\cong (\mathrm{S}^3,\rg_{\mathrm{round}})\,.
    \end{equation} 
    
    \subsubsection{Vacuum and adiabatic gauge field representatives}
    Because this example is so simple it allows us to illustrate our choice of representatives, discussed in section \ref{secreps}. Formula \eqref{bulkrepvac} becomes in this example
    \begin{equation}
    g_z(x)=g_{\psi,n}(x)=\exp\left(\lambda_+\frac{x}{R}\right)=\cos\left(\frac{\psi\,x}{R}\right)\,\mathbf{1}+i\sin\left(\frac{\psi\,x}{R}\right)\,n^I{\bm \sigma}_I\,,
    \end{equation}
    so that we find
    \begin{equation}
    \bar A(x;z)=\bar A(x;\psi,n)=g\,\partial_xg^{-1}dx=-2\psi\,n^I{\bm \sigma}_I\frac{dx}{R}\label{vacs}\,.
    \end{equation}
    Representatives of time dependent motion along these vacua are more involved, as discussed in section \ref{secadrep}, where we saw that the representatives get dressed by a particular local gauge transformation $h_z$ when extended into the bulk: $g_z=h_z\,\exp(\lambda_\uI z^I)$. This local factor is determined in terms of $\eta_z$, see \eqref{hform} and \eqref{etaform}.
    To find $\eta_z$ let us write 
    \begin{equation}
    \lambda(t,x)=\lambda_{\uI}(x)z^I(t)=\frac{2\psi(t)\,x}{R}\, n^I(t)\,T_I\,,
    \end{equation}
    and compute
    \begin{equation}
    e^{-\lambda}\frac{d}{dt} e^{\lambda}=i{\bm\sigma}_I\left[\frac{x\dot \psi}{R}\, n^I+\sin\left(\frac{x\psi}{R}\right)\left(\cos\left(\frac{x\psi}{R}\right)\,\dot n^I+\sin\left(\frac{x\psi}{R}\right)\,\epsilon_{IJK}\,n^J\,\dot n^K\right)\right]\,.
    \end{equation}
    Now this is an element of $\calsg$ and can be projected on $\calss$ via \eqref{1dproj}:
    \begin{equation}
    \sigma_z=\left(e^{-\lambda}\frac{d}{dt} e^{\lambda}\right)_\calss=i{\bm\sigma}_I\left[\frac{x}{R}\left(\dot\psi\, n^I+\sin\psi\cos\psi\,\dot n^I\right)+\sin^2\psi\,\epsilon_{IJK}\,n^J\,\dot n^K\right]\,.
    \end{equation}
    One now has all the ingredients to work out $\eta_z$:
    \begin{align}
\nonumber        \eta_z&=e^{\lambda}\left(e^{-\lambda}\frac{d}{dt} e^\lambda\right)_0e^{-\lambda}\\
\nonumber &=e^{\lambda}\left(e^{-\lambda}\frac{d}{dt} e^\lambda-\sigma_z\right)e^{-\lambda}\\
        &=\frac{i{\bm\sigma}_I}{4}\left[\left(x_-\sin(2x_+\psi)-x_+\sin(2x_-\psi)\right)\dot n^I+\left(x_-\cos(2x_+\psi)+x_+\cos(2x_-\psi)-1\right)\epsilon_{IJK}n^J\dot n^K\right]\label{eta1d}\,,
    \end{align}
    where
    \begin{equation}
    x_{\pm}=1\pm \frac{x}{R}\,.
    \end{equation}
    Note that one can easily check that $\eta_z$ is a local gauge parameter as it vanishes on the boundary, furthermore it is smooth all over the interval. It will be useful for the discussion below to already point out $\eta_z$ is orthogonal to $n^i$ and in particular proportional to $|\dot n|$. 
    
    In principle one can now compute $h_z$ via \eqref{hform}, but we were not able to do so in closed form. To finish this discussion, let us then present our adiabatic gauge field representatives:
    \begin{equation}
    A(t,x)=\bar A(x;z(t))=-2\psi(t)\,\frac{dx}{R}\,n^I(t)\,h_z(t,x)T_Ih_z^{-1}(t,x)+h_z(t,x)\partial_x h_z^{-1}(t,x)\,dx\label{adrep1d}\,.
    \end{equation}
    One can check that these configurations identically satisfy the Gauss constraint $D_i\dot A_i=0$, for any $(\psi(t),n(t))$, as they were defined to do.
    
     \subsubsection{Adiabatic solutions}
    Let us now specialize to the most interesting gauge fields, those corresponding to adiabatic solutions, or geodesics on $\calv$ and remember that everything we did so far is valid for any curve on the space of vacua. Because in this simple example $\calv$ turned out to be a symmetric space all geodesics will be one-parameter group orbits. In particular this will imply, as we mentioned in section \ref{sec-geodesics}, that the dressing factor $h_z$ will be trivial. Of course, because here $\calv$ is simply the round 3-sphere we will not need any of the general formalism. Let us focus on geodesics through the north-pole, i.e. $\psi=0$. The geodesics are then simply the great circles i.e. $\psi=\psi_0t$, $n=n_0$. So via the explicit formula \eqref{eta1d} we get confirmation of the fact that $h_z$ is trivial. Furthermore, via \eqref{adrep1d}, the adiabatic solutions are
    \begin{equation}
    A=-2\psi_0t\,\frac{dx}{R}\,n_0^I T_I\,.
    \end{equation}
    Writing $v^I=2\psi_0\,n_0^I$ the corresponding electric field $E=\dot A$ is of exactly the same type as (\ref{Efields},\,\ref{EfieldsYM}):
    \begin{equation}
    E=-\pd_x\Phi\qquad \Phi=v^I\,\frac{x}{R}\,T_I\,.
    \end{equation}
    Note that in this simple example the adiabatic approximation is exact, meaning that the above solution remains uncorrected even for large velocities, because in 1+1 dimensions the magnetic fields, and hence the potential, vanish identically so that all gauge fields are vacuum gauge fields.
    
    For this example the conserved charges are directly computed to be
    \begin{equation}
    Q_{\lambda_{\uI}}=P_I=\frac{2}{R}\mathbb{G}_{IJ}\,v^J\,.
    \end{equation}

    \section{Discussion and outlook}\label{discsec}
    In this work we showed how in the adiabatic limit the dynamics of Yang-Mills theory on a finite volume space with boundary contains purely electric solutions that are equivalent to geodesic motion on the space of vacua of the theory. Our analysis revealed this space of vacua to be the homogeneous space $\calv\cong \calg_0\backslash\calg/\calk$ equipped with a particular left invariant Riemannian metric. Furthermore we showed that these adiabatic solutions carry non-trivial multipole charges, that are `soft' in that they do not originate from matter sources. Here we will discuss some connections to the literature on closely related topics and mention how that might lead to interesting future research.
    
    Given that we were partially motivated by work on the infinite volume structure of gauge theories \cite{Strominger:2017zoo}, a first issue we should address is the large volume limit of our construction. We will leave a detailed study\footnote{Possibly along the lines of \cite{Andrade:2015fna}.} to future work and only mention some preliminary observations here. In the main text we found that the geodesic solutions are parameterized by a constant initial velocity $v$ and it follows from our computations that in $d$ spatial dimensions the corresponding charges scale as $Q\sim R^{d-2} v$ while the (kinetic) energy of the solutions scales as $K\sim R^{d-2}v^2$. This suggests that for $d\geq 3$ we could redefine $v=R^{2-d}w$ and keep $w$ fixed while sending $R\rightarrow \infty$, to ensure the solutions will keep non-vanishing, finite charges in this limit. There are now two interesting observations to be made. First note that in such a limit, the energy of the solutions will go to zero, which is reminiscent of the zero energy particles that appear in the soft theorems and the related soft modes generated by asymptotic symmetries \cite{He:2014laa,Strominger:2017zoo}. Secondly remark that this limit coincides with the adiabatic $v\rightarrow 0$ limit so that the solutions actually become exact in this large volume limit. The above arguments suggest that the space of vacua, its geometry and the associated geodesic solutions also exist in flat spacetime without boundary. It would be interesting to make this argument more robust and try and define these structures directly at infinite volume. We should point out that, as in \cite{Seraj:2016jxi}, our construction and the naive $R\rightarrow \infty$ limit mentioned above lead to asymptotic charges defined at spatial infinity, contrary to for example \cite{Strominger:2013lka,Barnich:2013sxa}  that define them at null infinity. Recently the connection between charges defined at spatial and null infinity has been investigated \cite{Campiglia:2017mua,Compere:2017knf,Troessaert:2017jcm,Prohazka:2017equ}. Note that in \cite{Mirbabayi:2016xvc} it was shown that the QED soft theorem also follows from the conservation of multipole charges defined at spatial infinity.
    
In our work there is a clear separation between the vacua, which are charge-less, and the `soft modes', i.e. the electric fields that correspond to slow motion along these vacua\footnote{That there might be a relation between soft photons and static electric fields is reminiscent of \cite{Barnich:2010bu,Muck:2015dea}.}, that carry non-zero charges. This is qualitatively different from what happens in gravitational theories, such as in AdS$_3$ or 4d Minkowski space, where the vacua themselves can carry non-trivial charges (see e.g. \cite{Banados:1998gg,Barnich:2010eb,Sheikh-Jabbari:2016unm,Compere:2016jwb,Oblak:2016eij}). This can be argued to be related to the central extension of the algebra of charges, which appears there but is absent in our setup. To see this, compute the variation of the charge $Q_{\l_1}$ under a global gauge transformation $\l_2$. Then in a centrally extended algebra, we find 
    \begin{align*}
    \de_{\l_2}Q_{\l_1}&=\{Q_{\l_1},Q_{\l_2}\}=Q_{[\l_1,\l_2]}+C(\l_1,\l_2)
    \end{align*}
    Now set the charges to be zero on a reference point `$o$'. Then the variation at that point is 
    \begin{align*}
    \de_{\l_2}Q_{\l_1}\Big\vert_o&=C(\l_1,\l_2)
    \end{align*}
    This implies that the central extension of the algebra makes the charges sensitive to the global gauge transformation. Note also that our vacua are smooth all over the space, while (coordinate) singularities usually appear in the gravitational context \cite{Strominger:2016wns, Compere:2016jwb}.
    
    There appears to be a `holographic' flavor to our work. It is interesting that in the adiabatic limit the (classical) dynamics of Yang-Mills theory reduces to a 1d sigma model which is purely determined in terms of boundary data. The electric fields corresponding to geodesic motion can be interpreted, via equation \eqref{Efields}, as arising from a boundary charge $v^{\ell m} Y_{\ell m}$ carried by the global/boundary gauge parameters. Our work also bears similarities to \cite{Donnelly:2014fua,Donnelly:2016auv,Geiller:2017xad}, with the global gauge parameters resembling the edge modes of that work.
    
    We should also emphasize that in this paper the analysis of the space of vacua has been mainly local, except for the example in section \ref{toyex}. In that example we found that the space of vacua is compact and has closed geodesics, which hints at interesting physics such as a discrete spectrum and non-trivial adiabatic phases\footnote{We thank B. Oblak for this last observation.}. Berry phases and their relation to infinite dimensional (asymptotic) symmetry algebras were recently explored in \cite{Oblak:2017ect}. Determining the precise global structure of the space of vacua and its geodesics for Yang-Mills theory in 3+1 dimensions, and studying the physical consequences thus seems an interesting future direction.
    
    In cosmology there has been work on the relation between certain adiabatic solutions and gauge symmetries \cite{Weinberg:2003sw, Hinterbichler:2013dpa, Mirbabayi:2016xvc}, it would be interesting to understand the relation to our work in more detail.
    
    Here we were only concerned with the geometry on the space of vacua, it would be interesting to extend this to the full configuration space. The geometry of the configuration space of Yang-Mills theory has been rather well studied, see e.g. \cite{Narasimhan:1979kf, Babelon:1980uj, Grabiak:1986si, Fuchs:1994zv, Orland:1996hm}, but in these works one restricts attention to compact spaces\footnote{More precisely, these works cover both compact Euclidean spaces and Lorentzian spacetimes of topology $\mathbb{R}\times\Sigma$ where the spatial hypersurface $\Sigma$ is either compact, or stringent boundary conditions are imposed such that no boundary term appears.}. Since our vacuum gauge fields are flat connections it might also be interesting to see if there are connections to the more mathematical work on spaces of such flat connections, see for example \cite{salamon1998notes}.
    
    Finally let us list some possible generalizations/extensions of this work that could prove interesting:
    \begin{enumerate}
    	\item Study of examples with multiple boundaries, which might be interpreted as defect insertions.  In the presence of an inner boundary for example, global gauge symmetries that would otherwise blow up in the bulk will be allowed. The corresponding adiabatic solutions  correspond to Coulomb like electric fields.
    	\item Inclusion of charged matter fields, which could lead to even richer vacuum manifolds. For example adding a charged scalar to Maxwell theory would make the action of the constant global gauge transformations non-trivial, so that they would not be quotiented out of the space of vacua.
    	\item Extension to the quantum regime, although it is not clear how much of the classical structure we uncovered will survive the strong quantum effects of Yang-Mills theory in the IR.
    	\item Study of adiabatic solutions and the space of vacua in gravitational theories. This is of particular interest as here it might be related to black hole entropy and the information paradox, see for example \cite{Hawking:2016msc,Afshar:2016uax}, although this is contested \cite{Bousso:2017dny, Bousso:2017rsx}.
    \end{enumerate}
   

 \section*{Acknowledgements}
 We dedicate this paper to the memory of John Freely and Maryam Mirakhani, and to the lives to come of Aylin and Honyeh.
 
 It is a pleasure to thank Glenn Barnich, Geoffrey Compère,  Marc Geiller, Blagoje Oblak, Peter Orland and Andy Royston for discussions and useful suggestions. In particular we would like to thank Emine S Kutluk for her contributions in the early stages of this work.
 
 DVdB was partially supported by the Bo\u{g}azi\c{c}i University Research Fund under grant number 17B03P1.
 AS would like to thank the ICTP network scheme NT-04, Bonyad Melli Nokhbegan (BMN) and SarAmadan club of Iran for their partial support, and the Istanbul Center for Mathematical Sciences (IMBM) for the hospitality.
    
    \appendix
    \section{Technicalities on gauge transformations}\label{diffeqapp}
     In this appendix we spell out some details on the derivation of certain properties of subclasses of infinitesimal gauge transformations that are used in the main text.

     \paragraph{Property 1)} If $D^2\gamma=0$ and $\left.\gamma\right|_{\partial M}=0$ then $\gamma=0$\,.

     This follows from the observation that
     \begin{equation}
     \int_M d^3 x\,\Tr\, D_i \gamma D_i\gamma=-\int_M d^3 x\,\Tr\, \gamma D^2\gamma\,+\,\oint_{\partial M}\!\!d\Sigma^i\, \Tr\, \gamma D_i\gamma=0\,.
     \end{equation}
     However, since the left hand side is manifestly positive definite it follows that $D\gamma=0$. This set of first order differential equations has a unique solution given the boundary condition $\left.\gamma\right|_{\partial M}=0$, which is $\gamma=0$.

      \paragraph{Property 2)} One has the vector space decomposition $\calsg=\calsg_0\oplus\calss$\,.
   
     In section \ref{secgauss} we showed that $\calsg_0$ is orthogonal to $\calss$. However we should be careful as the inner product used is degenerate, so that there could be a subspace of vectors that are in both $\calsg_0$ and $\calss$. But 1) above is equivalent to $\calsg_0\cap\calss=\{0\}$ so that $\calsg_0$ complements $\calss$.
     
      \paragraph{Property 3)}For each $\gamma\in\calsg$ there is a unique $\sigma\in\calss$ such that $\left.\gamma\right|_{\partial M}=\left.\sigma\right|_{\partial M}$\,.
     
     This is equivalent to there being a unique solution of $D^2\sigma=0$ for given boundary values, which is a Dirichlet problem. Assuming existence, uniqueness follows from that fact that given two solutions $\sigma_1$ and $\sigma_2$, their difference satisfies the conditions of 1), and hence is zero.

     \paragraph{Property 4)} $\ker \Dperp=\ker D$
      
     First note that both operators have a different domain, $\cals$ and $\calsg$, and image, $\calss$ and $\mathbb{R}^3\otimes \calsg$, respectively. 
     
     It follows rather directly that $\ker D\subset \ker \Dperp$. First observe that when $D\kappa=0$ it follows that $D\kappa_0=-D\kappa_\calss$ and thus $D^2\kappa_0=0$. But then via 1) $\kappa_0=0$ which means that $\kappa\in\calss$ and $D\kappa_\calss=0$. It then directly follows that $\left.(\Dperp \kappa)\right|_{\partial M}=\left(n^iD_{i}\kappa_\calss)\right|_{\partial M}=0$ and because $\Dperp\kappa\in\calss$ it follows via 3) that $\Dperp\kappa=0$. 
     
     To show the reverse, i.e. that $\ker \Dperp\subset \ker D$  is also true, consider an element $\kappa\in \ker \Dperp$ which implies both $D^2\kappa=0$ and $\Dperp\kappa=0$. But then observe that 
     \begin{equation}
     \int_M\!d^3x\,
    \Tr\,D_{i}\kappa D_{i}\kappa= \oint_{\partial M}\!\!d\Sigma \,\Tr\,\kappa\Dperp\kappa=0\,.
     \end{equation}
     Because the left hand side is positive definite it follows that indeed also $D\kappa=0$.
     
      \paragraph{Property 5)}$[\kappa, \Dperp\sigma]_*=\Dperp[\kappa,\sigma]_*$ for all $\sigma\in\calss$ and $\kappa\in\calsk$. 
     
     Because both sides of the equality above are valued in $\calss$ and the property 3) it is enough to show equality on the boundary:
     \begin{equation}
     \left.n^i([\kappa, D_i\sigma])\right|_{\partial M}=n^i\left.(D_i[\kappa,\sigma])\right|_{\partial M}\,.
     \end{equation}
     However, this immediately follows from the Liebnitz property of the gauge covariant derivative, i.e. $D[\gamma_1,\gamma_2]=[D\gamma_1,\gamma_2]+[\gamma_1,D\gamma_2]$, and that $D\kappa=0$ as $\calsk=\ker D$.

      \paragraph{Property 6)}Any element $g\in\calg$ can  be written on the boundary as
     	\begin{equation}
     	\left.g\right|_{\partial M}=\left.\exp(\lambda)\right|_{\partial M}\left.k\right|_{\partial M}\,,\qquad\mbox{where}\quad\lambda\in\calsm\ \ \mbox{and}\ \ k\in\calk\,.
     	\end{equation} 
     
      This is rather straightforward to derive:
      \begin{eqnarray}
\nonumber      \left.g\right|_{\partial M}&=&\left.\exp(\gamma)\right|_{\partial M}\\
\nonumber      &=&\left.\exp(\gamma_\calss)\right|_{\partial M}\\
\nonumber      &=&\left.\exp(\gamma_{\calsk}+\gamma_\calsm)\right|_{\partial M}\\
      &=&\left.\exp(\lambda)\exp(\gamma_{\calsk})\right|_{\partial M}\,.
      \end{eqnarray}
      That $\lambda\in\calsm$ follows from the Baker-Campbell-Haussdorf formula, which expresses $\lambda$ as an expansion in commutators of the form $[\gamma_{\calsk},\ldots,[\gamma_{\calsk},\gamma_{\calsm}]_*]_*$\,, and the property \eqref{reductive property} of reductive homogeneous spaces i.e. $[\calsk,\calsm]_*\subset\calsm$.

    It should be noticed that the uniqueness of the exponential representation is only guaranteed in a patch around $z=0$. Indeed a global choice of coset representatives typically does not exist.    
    \section{Some facts on homogeneous spaces}\label{sec-homogeneous spaces}
    In this appendix we collect some definitions, concepts and results of the mathematical theory of cosets and homogeneous spaces. We will make many statements without proofs or arguments, which can be found in various standard references on the subject. Note that much of the literature deals with the special cases of symmetric or normal reductive spaces but the spaces we discuss will be more general. A pedagogic reference that deals with the general reductive case is \cite{arvanitogeorgos2003introduction} of which we will follow certain parts quite closely. A physics paper containing a nice review and further references is \cite{Castellani:1983tb}. In the second subsection on Riemannian homogeneous spaces we also use some details that can be found in the original work \cite{KowVa}. Although the discussion in all these references is for finite dimensional manifolds it appears, heuristically, that the relevant facts generalize to the particular infinite dimensional case we are dealing with in the main text.

    \subsection{Homogeneous spaces}\label{aphom}
    \paragraph{Basics on cosets}
    
    \begin{itemize}
    	\item Consider a subgroup $K$ of a group $G$. One can then define an equivalence relation between two elements $g_1, g_2\in G$ by $g_1\sim g_2$ iff $g_2^{-1}g_1 \in K$. The equivalence class containing an element $g\in G$ is denoted as $gK$ and is called a {\it left coset}. The set of left cosets is denoted by $G/K$. 
    	\item Although in general the coset space $G/K$ does not form a group, there exists a natural left $G$ action on it: 
    	\begin{equation}
    	g_1\cdot (gK)=(g_1g)K\,.\label{groupaction}
    	\end{equation}
    	\item In complete parallel {\it right cosets} $Kg$ can be defined leading to $K\backslash G$, which has a natural right $G$ action on it. 
    	\item If the left cosets are equal to the right cosets, i.e. $gK=Kg$ for all $g\in G$, the subgroup $K$ is called {\it normal} and the coset space $G/K=K\backslash G$ has a natural group structure on it via $g_1 K g_2 K=(g_1 g_2)K$, which is called the {\it quotient group}. 
    	\item Given a homomorphism $\phi : G\rightarrow H$ it follows that $\ker\phi$ is a normal subgroup of $G$ and $\phi(G)$ is a subgroup of $H$. Moreover, the following isomorphism holds:
    	\begin{equation}
    	G/\ker \phi\cong \phi(G)\,.
    	\end{equation}
    	This fact goes under the name of the first isomorphism theorem and is often useful in identifying quotient groups.  
    \end{itemize}

    \paragraph{The differentiable case}
    \begin{itemize}
    	\item When $G$ is a Lie group and $K$ is a closed subgroup, the coset space $G/K$ naturally is a manifold with a smooth left $G$ action, and is called a {\it homogeneous space}.
    	
    	\item  Consider a manifold $M$ with a smooth left $G$ action that is {\it transitive}, which means that for any two points $p_1, p_2\in M$ there exists a $g\in G$ such that $g\cdot p_1=p_2$. One can define the {\it isotropy subgroup} $G_p$ of a point $p \in M$ as the subgroup that leaves the point $p$ invariant, i.e. $G_p=\{g\in G|\;g\cdot p=p\}$. It follows that the $G_p$'s are isomorphic for all $p$ and furthermore closed, so that there exists a natural diffeomorphism :
    	\begin{equation}
    	M\cong G/K\,,
    	\end{equation}
    	where $K\cong G_p$\,. In practice this diffeomorphism amounts to picking an arbitrary reference point $p_o\in M$ and mapping $gK$ to $p=g\cdot p_o$. 
    	\item Concretely one can choose local coordinates $z^a$ on $G/K$ through a choice of unique representative $g_z$ out of each equivalence class.
    	\item Denoting the Lie algebras of $G$ and $K$ with $\mathfrak{g}$ and $\mathfrak{k}$ respectively, one defines the quotient $\mathfrak{g}/\mathfrak{k}$ as the set of equivalence classes $X+\mathfrak{k}=Y+\mathfrak{k}$ iff $ X-Y \in \mathfrak{k}$.
    	\item In general the vector space $\mathfrak{g}/\mathfrak{k}$ does not have an algebra structure, except when $\mathfrak{k}$ is a {\it Lie algebra ideal} of $\mathfrak{g}$, which means $[X,Y]\in \mathfrak{k}\,,\ \forall X\in \mathfrak{g}$, $Y\in \mathfrak{k}$. This is the infinitesimal analog of $K$ being a normal subgroup. In that case $\mathfrak{g}/\mathfrak{k}$ comes naturally  equipped with the Lie bracket $[X+\mathfrak{k},Y+\mathfrak{k}]_{\mathfrak{g}/\mathfrak{k}}=[X,Y]+\mathfrak{k}$. As might be expected $\mathfrak{g}/\mathfrak{k}$ is the Lie algebra of $G/K$.
    	
    	\item Even when $\mathfrak{k}$ is not an ideal, one can still identify the quotient vector space with the tangent space of the homogenous space at the reference point: $\mathfrak{g}/\mathfrak{k}\cong T_{eK}G/K\cong T_oM$.
    	
    \end{itemize}
    \subsection{Riemannian homogeneous spaces}\label{apmath}
    
    \paragraph{Basic definitions}
    \begin{itemize}
    	\item A {\it reductive homogeneous space} is a coset space $M=G/K$ such that there exists an $\mathrm{Ad}(K)$ invariant decomposition $\mathfrak{g}=\mathfrak{k}\oplus \mathfrak{m}$. This is equivalent to the conditions $[\mathfrak{k},\mathfrak{k}]\subset\mathfrak{k}$ and $[\mathfrak{k},\mathfrak{m}]\subset\mathfrak{m}$. In this case, we can identify $\mathfrak{m}\cong \mathfrak{g}/\mathfrak{k}\cong T_{eK}(G/K)\cong T_oM$.
    	\item A {\it Riemannian homogeneous space} is a Riemannian manifold $(M, \rg)$ with a transitive action of a group of isometries $G$. This implies we can identify $M\cong G/K$, with $K$ the group of isometries that leave a point invariant.
    \end{itemize}

    \paragraph{Properties of Riemannian homogeneous spaces} 
    \begin{itemize}
    	\item A Riemannian homogeneous space is reductive.
    	\item If $G$ is a compact group, $G/K$ is reductive, as one can take $\mathfrak{m}$ to be the orthogonal complement of $\mathfrak{k}$, with respect to the Killing form.
    	\item In case $K$ is compact, any metric invariant under the left $G$-action can be written as
    	\begin{equation}
    	\rg_{ab}(z)=\langle e_a(z),e_b(z)\rangle=(e_a(z),\Dperp\, e_b(z))=\Dperp_{\ua\ub}\,e_a^\ua(z)\,e_b^\ub(z)\,,\label{appmetric}
    	\end{equation}
    	Here we have chosen coordinates $z^a$ and a set of representatives $g_z$. The 'vielbein' $e_a=(g_z^{-1}\partial_a g_z)_\mathfrak{m}$ is the projection on $\mathfrak{m}$ of the left invariant one-form $\theta=g_z^{-1}\rd g_z$. The $z$-independent operator $\Dperp$ on $\mathfrak{m}$ is symmetric with respect to the bi-invariant scalar product $(\cdot,\cdot)$ on $\mathfrak{g}$ and furthermore $\Ad(K)$-equivariant. By choosing a basis $T_\ua$ for $\mathfrak{m}$ these properties are expressed in terms of the matrix elements $\Dperp_{\ua\ub}=(T_\ua,\Dperp\,T_\ub)$ as
    	\begin{equation}
    	\mathbb{D}_{\ua\ub}=\mathbb{D}_{\ub\ua}\qquad\mbox{and}\qquad S_{\ua}{}^{\uc}(k)S_{\ub}{}^{\ud}(k)\mathbb{D}_{\uc\ud}=\mathbb{D}_{\ua\ub}\quad\forall k\in K\,,\label{Dtransf}
    	\end{equation}
    	where
    	\begin{equation}
    	\Ad(k)T_\ua=k^{-1}T_\ua k=S_{\ua}{}^{\ub}(k)T_\ub\,.
    	\end{equation}
    	\item Note that the metric $\rg$ above is independent of the choice of representative. This follows because under a change of representative $g_z\rightarrow g_z k$ and $e_a\rightarrow k^{-1}e_ak$, so that due to the equivariance of $\Dperp$ and invariance of $(\cdot,\cdot)$ the metric $\rg$ remains invariant.
    	\item The left invariance of $\rg$ follows by the same argument, as after the left action by a constant element $g_1\in G$ the new representative is $g_z'=g_1 g_z k^{-1}$ and $e_a'=ke_ak^{-1}$. Infinitesimally this action associates to any element $X\in \mathfrak{g}$ a vector field
    	\begin{equation}
    	\xi^a[X]=e^a_\ua\,(g_z^{-1}Xg_z)^\ua\,,
    	\end{equation}
    	with $e^a_\ua$ the inverse vielbein and  $(g_z^{-1}Xg_z)_\mathfrak{m}=(g_z^{-1}Xg_z)^\ua\,T_\ua$. One can check by direct calculation that these vector fields satisfy the Killing equation and that
    	\begin{equation}
    	\left[\,\xi[X],\xi[Y]\,\right]^a=-\xi^a\left[\,[X,Y]\,\right]\,.
    	\end{equation}
    	
    \end{itemize}
    
    \paragraph{Some special subclasses}
    \begin{itemize}
    	\item A {\it symmetric space} is a reductive homogeneous space such that $[\mathfrak{m},\mathfrak{m}]\subset\mathfrak{k}$\,. 
    	\item A {\it naturally reductive space} is a Riemannian homogeneous space for which 
    	\begin{equation}
    	[T_\ua,T_\ub]^\uc\,\Dperp_{\uc\ud}=\Dperp_{\ua\uc}\,[T_\ub,T_\ud]^\uc\,,
    	\end{equation}
    	where $[T_\ua,T_\ub]_\mathfrak{m}=[T_\ua,T_\ub]^\uc\,T_\uc$, or in terms of the structure constants $[T_\ua,T_\ub]^\uc=f_{\ua\ub}{}^\uc$\,.
    	\item A {\it g.o.\! space} is a Riemannian homogeneous space for which all geodesics are an orbit of a one-parameter subgroup of isometries, i.e. $\gamma(t)=\exp(tX)\cdot p$\,, for some $X\in\mathfrak{g}$.

    	\noindent The above definitions imply that
    	\begin{equation*}
    	\mbox{symmetric }\subset \mbox{ naturally reductive }\subset\mbox{ g.o. }\subset\mbox{ reductive}\label{classif}
    	\end{equation*}
    	\item There are also {\it normal reductive spaces}, defined as those for which $\Dperp=\mathbf{1}$, or in other words $\langle\cdot,\cdot\rangle=\left(\cdot,\cdot\right)$. One can check that
    	\begin{equation}
    	\mbox{normal reductive}\subset \mbox{ naturally reductive }\,.
    	\end{equation}
    \end{itemize}

    \paragraph{Orbits that are geodesics} (Proposition 2.1 of \cite{KowVa}) An orbit $\exp(t X)\cdot o$ is a geodesic iff
    \begin{equation}
    \langle[X,Y]_\mathfrak{m},X_\mathfrak{m}\rangle=0\qquad \forall Y\in\mathfrak{g}\label{geovec}\,.
    \end{equation}
    
    In this case $X$ is called a {\it geodesic vector}. We provide a physicists proof of this proposition in (\ref{startphys}-\ref{endphys}). As pointed out in \cite{arvanitogeorgos2003introduction} there are a number of conditions that are sufficient as well. The one that will be particularly relevant for us is that if $[X_\mathfrak{m},\Dperp X_\mathfrak{m}]=0$ then the orbit generated by $X$ is a geodesic.
    
    \paragraph{A criterion to be g.o.} (Proposition 2.6 of \cite{KowVa}) If a Riemannian homogeneous space $G/K$ is g.o. then for all $X\in\mathfrak{m}$ there exists a $Z\in\mathfrak{k}$ such that for all $Y\in\mathfrak{m}$
    \begin{equation}
    \langle[X+Z,Y]_\mathfrak{m},X\rangle=0\label{crit}\,.
    \end{equation}
    
    \paragraph{Proof that the space of vacua is in general not g.o.}
    We can now apply the criterion \eqref{crit} to the gauge theory context discussed in the main text. In the notation used in the main text, the criterion above states that for all $\lambda_1\in\calsm$ there should exist a $\kappa\in\calsk$ such that for all $\lambda_2\in\calsm$ the quantity  $\langle[\lambda_1+\kappa,\lambda_2]_\calss,\lambda_1\rangle$ should vanish. But this is not the case for non-abelian Yang-Mills theory in $d+1$ dimensions when $d\geq1$.  Consider the case $d=3$, for which we discussed the split $\calss=\calsk\oplus\calsm$ in section \ref{4dYMsplit}. Now take for example
    \begin{equation*}
    \lambda_1=\left(\frac{r}{R}\right)T_{I_1} Y_{10}+\left(\frac{r}{R}\right)^2T_{I_2} Y_{20}\,,\quad \lambda_2=\left(\frac{r}{R}\right)^3[T_{I_1},T_{I_2}] Y_{30}\,,
    \end{equation*}
    for which one computes that, independent of $\kappa$, 
    \begin{equation*}
    \langle[\lambda_1+\kappa,\lambda_2]_{*\,\calsm},\lambda_1\rangle=\left([\lambda_1+\kappa,\lambda_2]_*,\Dperp\lambda_1\right)=-\frac{3}{2} \sqrt{\frac{3}{35 \pi }}\Tr\, [T_{I_1},T_{I_2}][T_{I_1},T_{I_2}]\,.
    \end{equation*}
    Because the gauge algebra is semi-simple, choosing $T_{I_1}$ and $T_{I_2}$ to be non-commuting suffices to make the above non-vanishing, which implies that the space of vacua $\calv$ is not g.o. .

    \section{Charges from covariant phase space}\label{app-charges}
    In this appendix, we briefly discuss the construction of charges in the covariant phase space formulation \cite{Lee:1990nz,Wald:1993nt,Iyer:1994ys,Barnich:2001jy} (See \cite{,Seraj:2016cym} for a review).
    
    Suppose that a gauge theory is given with dynamical fields collectiely denoted by $\psi$ and Lagrangian $\bL[\psi]$ as a top form. Varying the Lagrangian gives the Euler-Lagrange equations plus a total derivative
    \begin{align}
    \de\bL&=\bE \de \psi+ d\bTheta(\de\psi)\,.
    \end{align}
    The codimension-1 form $\bTheta$ is called the presymplectic potential from which the presymplectic current $\bomega$ is constructed by taking another antisymmetric variation
    \begin{align}
    \bomega(\psi,\de_1\psi,\de_2\psi)&=\de_1\bTheta(\de_2\psi)-\de_2\bTheta(\de_1\psi)\,.
    \end{align}
    The pre-symplectic form $\Omega(\psi,\de_1\psi,\de_2\psi)$ on the phase space is then defined as the integral of $\bomega$ over  a Cauchy surface $\Sigma$
    \begin{align}\label{symplectic form def}
    \Omega(\psi,\de_1\psi,\de_2\psi)&=\int_\Sigma \bomega(\psi,\de_1\psi,\de_2\psi)\,.
    \end{align}
    This quantity defines a closed  two form on the space of field configurations. However, it's degeneracies prevents its identification with a symplectic form. This is why it is called a \textit{pre}-symplectic form. Indeed, the problem arises as the space of field configurations is too large. Therefore, the covariant phase space is defined as the symplectic quotient of the space of field configurations by the degeneracies of the presymplectic form \cite{Woodhouse:1980pa,Lee:1990nz}.
    
    The Hamiltonian generator of a symmetry transformation $\psi\to \psi+\de_\gamma \psi$ (in the setting of this paper a gauge transformation $\gamma\in\calsg$) is then given by
    \begin{align}\label{var H def}
    \de H_\gamma &=\Omega(\psi,\de\psi,\de_\gamma\psi)\,.
    \end{align}
    As $\gamma(x)$ is the parameter of a \textit{local} symmetry transformation, it turns out that the symplectic current is on-shell exact \cite{Iyer:1994ys,Barnich:2001jy}, i.e. 
    \begin{align}\label{theorem-charges}
    \bomega(\psi,\de\psi,\de_\gamma\psi)&=d\,\bk_\gamma (\psi,\de\psi)\qquad\mbox{on-shell}\,.
    \end{align}
    Accordingly, the charge $\de Q_\gamma$ defined as the on-shell value of the Hamiltonian becomes a surface integral
    \begin{align}\label{def-Hamiltonian}
    \de H_\gamma &=\oint_{\pd\Sigma}\bk_\gamma (\psi,\de\psi)\,.
    \end{align}
    
    Now let us apply the above construction to pure Yang-Mills theory with dynamical fields $A_\mu$ and Lagrangian\eqref{actionYM}. We compute presymplectic potential and current in their dual form
    \begin{align}
    \bTheta&=\star(\theta_\mu dx^\mu),\qquad \bomega=\star(\omega_\mu dx^\mu)\,.
    \end{align}
    It can be checked that 
    \begin{align}
    \theta^\mu(\de\psi)&=\Tr\, F^\mn \de A_\nu\,,
    \end{align}
    and accordingly
    \begin{align}\label{symplectic form EM}
    \omega^\mu(A,\de_1A,\de_2A) &=\Tr\, \de_1 F^{\mn}\de_2 A_\nu-(1\leftrightarrow 2)\,.
    \end{align}
    The charge variation can then be computed as
    \begin{align}
    \de Q_\gamma&=\Omega(\psi,\de\psi,\de_\gamma\psi)=\int_{\Sigma}\Tr\, \de F^{\mn}\pd_\nu \gamma=\oint_{\pd\Sigma} d\Sigma_\mn\, \Tr \de F^\mn \gamma(x)\,.
    \end{align}
    This expression is clearly integrable and hence we find the final expression for the charges:
    \begin{align}\label{charges-YM}
    \boxed{Q_\gamma=\oint_{\pd\Sigma} d\Sigma_\mn\, \Tr\,F^\mn \gamma(x)\,.}
    \end{align}
    The Poisson bracket of two charges can also be computed using the symplectic structure 
    \begin{align}\label{Poisson bracket}
            \{Q_{\gamma_1},Q_{\gamma_2}\}&=\Omega(A,\de_{\gamma_1} A,\de_{\gamma_2}A)=\de_{\gamma_2} Q_{\gamma_1}
    \end{align}

    \bibliographystyle{JHEP}
    \bibliography{vac_lit}

\providecommand{\href}[2]{#2}\begingroup\raggedright\begin{thebibliography}{10}

\bibitem{Strominger:2017zoo}
A.~Strominger, \emph{{Lectures on the Infrared Structure of Gravity and Gauge
  Theory}},  \href{https://arxiv.org/abs/1703.05448}{{\ttfamily 1703.05448}}.

\bibitem{Seraj:2016jxi}
A.~Seraj, \emph{{Multipole charge conservation and implications on
  electromagnetic radiation}},
  \href{https://doi.org/10.1007/JHEP06(2017)080}{\emph{JHEP} {\bfseries 06}
  (2017) 080}, [\href{https://arxiv.org/abs/1610.02870}{{\ttfamily
  1610.02870}}].

\bibitem{Strominger:2013lka}
A.~Strominger, \emph{{Asymptotic Symmetries of Yang-Mills Theory}},
  \href{https://doi.org/10.1007/JHEP07(2014)151}{\emph{JHEP} {\bfseries 07}
  (2014) 151}, [\href{https://arxiv.org/abs/1308.0589}{{\ttfamily 1308.0589}}].

\bibitem{Barnich:2013sxa}
G.~Barnich and P.-H. Lambert, \emph{{Einstein-Yang-Mills theory: Asymptotic
  symmetries}}, \href{https://doi.org/10.1103/PhysRevD.88.103006}{\emph{Phys.
  Rev.} {\bfseries D88} (2013) 103006},
  [\href{https://arxiv.org/abs/1310.2698}{{\ttfamily 1310.2698}}].

\bibitem{Campiglia:2017mua}
M.~Campiglia and R.~Eyheralde, \emph{{Asymptotic $U(1)$ charges at spatial
  infinity}},  \href{https://arxiv.org/abs/1703.07884}{{\ttfamily 1703.07884}}.

\bibitem{Weinberg:2006rq}
E.~J. Weinberg and P.~Yi, \emph{{Magnetic Monopole Dynamics, Supersymmetry, and
  Duality}}, \href{https://doi.org/10.1016/j.physrep.2006.11.002}{\emph{Phys.
  Rept.} {\bfseries 438} (2007) 65--236},
  [\href{https://arxiv.org/abs/hep-th/0609055}{{\ttfamily hep-th/0609055}}].

\bibitem{Weinberg:2012pjx}
E.~J. Weinberg, \emph{{Classical solutions in quantum field theory}}.
\newblock Cambridge Monographs on Mathematical Physics. Cambridge University
  Press, 2015.

\bibitem{Atiyah:1988jp}
M.~F. Atiyah and N.~J. Hitchin, \emph{{The Geometry and Dynamics of Magnetic
  Monopoles. M.B. Porter Lectures}}.
\newblock 1988.

\bibitem{Manton:1981mp}
N.~S. Manton, \emph{{A Remark on the Scattering of BPS Monopoles}},
  \href{https://doi.org/10.1016/0370-2693(82)90950-9}{\emph{Phys. Lett.}
  {\bfseries B110} (1982) 54--56}.

\bibitem{Julia:1975ff}
B.~Julia and A.~Zee, \emph{{Poles with Both Magnetic and Electric Charges in
  Nonabelian Gauge Theory}},
  \href{https://doi.org/10.1103/PhysRevD.11.2227}{\emph{Phys. Rev.} {\bfseries
  D11} (1975) 2227--2232}.

\bibitem{Lechtenfeld:2015uka}
O.~Lechtenfeld and A.~D. Popov, \emph{{Yang–Mills moduli space in the
  adiabatic limit}},
  \href{https://doi.org/10.1088/1751-8113/48/42/425401}{\emph{J. Phys.}
  {\bfseries A48} (2015) 425401},
  [\href{https://arxiv.org/abs/1505.05448}{{\ttfamily 1505.05448}}].

\bibitem{Popov:2015wsa}
A.~D. Popov, \emph{{Loop groups in Yang–Mills theory}},
  \href{https://doi.org/10.1016/j.physletb.2015.07.041}{\emph{Phys. Lett.}
  {\bfseries B748} (2015) 439--442},
  [\href{https://arxiv.org/abs/1505.06634}{{\ttfamily 1505.06634}}].

\bibitem{Lechtenfeld:2015waa}
O.~Lechtenfeld and A.~D. Popov, \emph{{Supermembrane limit of Yang-Mills
  theory}}, \href{https://doi.org/10.1063/1.4942186}{\emph{J. Math. Phys.}
  {\bfseries 57} (2016) 023520},
  [\href{https://arxiv.org/abs/1508.06325}{{\ttfamily 1508.06325}}].

\bibitem{Lechtenfeld:2016sgc}
O.~Lechtenfeld and A.~D. Popov, \emph{{Superstring limit of Yang–Mills
  theories}}, \href{https://doi.org/10.1016/j.physletb.2016.09.032}{\emph{Phys.
  Lett.} {\bfseries B762} (2016) 309--314},
  [\href{https://arxiv.org/abs/1608.05331}{{\ttfamily 1608.05331}}].

\bibitem{arnol2013mathematical}
V.~I. Arnol'd, \emph{Mathematical methods of classical mechanics}, vol.~60.
\newblock Springer Science \& Business Media, 2013.

\bibitem{Stuart:2007zz}
D.~M.~A. Stuart, \emph{{Analysis of the adiabatic limit for solitons in
  classical field theory}},
  \href{https://doi.org/10.1098/rspa.2007.0130}{\emph{Proc. Roy. Soc. Lond.}
  {\bfseries A463} (2007) 2753--2781}.

\bibitem{Barnich:2010eb}
G.~Barnich and C.~Troessaert, \emph{{Aspects of the BMS/CFT correspondence}},
  \href{https://doi.org/10.1007/JHEP05(2010)062}{\emph{JHEP} {\bfseries 05}
  (2010) 062}, [\href{https://arxiv.org/abs/1001.1541}{{\ttfamily 1001.1541}}].

\bibitem{Compere:2015knw}
G.~Compère, P.~Mao, A.~Seraj and M.~M. Sheikh-Jabbari, \emph{{Symplectic and
  Killing symmetries of AdS$_{3}$ gravity: holographic vs boundary gravitons}},
  \href{https://doi.org/10.1007/JHEP01(2016)080}{\emph{JHEP} {\bfseries 01}
  (2016) 080}, [\href{https://arxiv.org/abs/1511.06079}{{\ttfamily
  1511.06079}}].

\bibitem{Castellani:1983tb}
L.~Castellani, L.~J. Romans and N.~P. Warner, \emph{{Symmetries of Coset Spaces
  and {Kaluza-Klein} Supergravity}},
  \href{https://doi.org/10.1016/0003-4916(84)90066-6}{\emph{Annals Phys.}
  {\bfseries 157} (1984) 394}.

\bibitem{Stein}
H.~H. Homeier and E.~O. Steinborn, \emph{{Some properties of the coupling
  coefficients of real spherical harmonics and their relation to Gaunt
  coefficients}}, {\emph{Journal of Molecular Structure} {\bfseries 368} (1996)
  31--37}.

\bibitem{arvanitogeorgos2003introduction}
A.~Arvanitoge{\=o}rgos, \emph{An introduction to Lie groups and the geometry of
  homogeneous spaces}, vol.~22.
\newblock American Mathematical Soc., 2003.

\bibitem{Andrade:2015fna}
T.~Andrade and D.~Marolf, \emph{{Asymptotic Symmetries from finite boxes}},
  \href{https://doi.org/10.1088/0264-9381/33/1/015013}{\emph{Class. Quant.
  Grav.} {\bfseries 33} (2016) 015013},
  [\href{https://arxiv.org/abs/1508.02515}{{\ttfamily 1508.02515}}].

\bibitem{He:2014laa}
T.~He, V.~Lysov, P.~Mitra and A.~Strominger, \emph{{BMS supertranslations and
  Weinberg’s soft graviton theorem}},
  \href{https://doi.org/10.1007/JHEP05(2015)151}{\emph{JHEP} {\bfseries 05}
  (2015) 151}, [\href{https://arxiv.org/abs/1401.7026}{{\ttfamily 1401.7026}}].

\bibitem{Compere:2017knf}
G.~Compère and A.~Fiorucci, \emph{{Asymptotically flat spacetimes with BMS$_3$
  symmetry}},  \href{https://arxiv.org/abs/1705.06217}{{\ttfamily 1705.06217}}.

\bibitem{Troessaert:2017jcm}
C.~Troessaert, \emph{{The BMS4 algebra at spatial infinity}},
  \href{https://arxiv.org/abs/1704.06223}{{\ttfamily 1704.06223}}.

\bibitem{Prohazka:2017equ}
S.~Prohazka, J.~Salzer and F.~Schöller, \emph{{Linking Past and Future Null
  Infinity in Three Dimensions}},
  \href{https://doi.org/10.1103/PhysRevD.95.086011}{\emph{Phys. Rev.}
  {\bfseries D95} (2017) 086011},
  [\href{https://arxiv.org/abs/1701.06573}{{\ttfamily 1701.06573}}].

\bibitem{Mirbabayi:2016xvc}
M.~Mirbabayi and M.~Simonović, \emph{{Weinberg Soft Theorems from Weinberg
  Adiabatic Modes}},  \href{https://arxiv.org/abs/1602.05196}{{\ttfamily
  1602.05196}}.

\bibitem{Barnich:2010bu}
G.~Barnich, \emph{{The Coulomb solution as a coherent state of unphysical
  photons}}, \href{https://doi.org/10.1007/s10714-010-0984-6}{\emph{Gen. Rel.
  Grav.} {\bfseries 43} (2011) 2527--2530},
  [\href{https://arxiv.org/abs/1001.1387}{{\ttfamily 1001.1387}}].

\bibitem{Muck:2015dea}
W.~Mück, \emph{{Photons in a Ball}},
  \href{https://doi.org/10.1140/epjc/s10052-015-3811-0}{\emph{Eur. Phys. J.}
  {\bfseries C75} (2015) 585},
  [\href{https://arxiv.org/abs/1510.04490}{{\ttfamily 1510.04490}}].

\bibitem{Banados:1998gg}
M.~Banados, \emph{{Three-dimensional quantum geometry and black holes}},
  \href{https://arxiv.org/abs/hep-th/9901148}{{\ttfamily hep-th/9901148}}.

\bibitem{Sheikh-Jabbari:2016unm}
M.~M. Sheikh-Jabbari and H.~Yavartanoo, \emph{{On 3d bulk geometry of Virasoro
  coadjoint orbits: orbit invariant charges and Virasoro hair on locally
  AdS$_3$ geometries}},
  \href{https://doi.org/10.1140/epjc/s10052-016-4326-z}{\emph{Eur. Phys. J.}
  {\bfseries C76} (2016) 493},
  [\href{https://arxiv.org/abs/1603.05272}{{\ttfamily 1603.05272}}].

\bibitem{Compere:2016jwb}
G.~Compère and J.~Long, \emph{{Vacua of the gravitational field}},
  \href{https://doi.org/10.1007/JHEP07(2016)137}{\emph{JHEP} {\bfseries 07}
  (2016) 137}, [\href{https://arxiv.org/abs/1601.04958}{{\ttfamily
  1601.04958}}].

\bibitem{Oblak:2016eij}
B.~Oblak, \emph{{BMS Particles in Three Dimensions}},
  \href{https://arxiv.org/abs/1610.08526}{{\ttfamily 1610.08526}}.

\bibitem{Strominger:2016wns}
A.~Strominger and A.~Zhiboedov, \emph{{Superrotations and Black Hole Pair
  Creation}},  \href{https://arxiv.org/abs/1610.00639}{{\ttfamily 1610.00639}}.

\bibitem{Donnelly:2014fua}
W.~Donnelly and A.~C. Wall, \emph{{Entanglement entropy of electromagnetic edge
  modes}}, \href{https://doi.org/10.1103/PhysRevLett.114.111603}{\emph{Phys.
  Rev. Lett.} {\bfseries 114} (2015) 111603},
  [\href{https://arxiv.org/abs/1412.1895}{{\ttfamily 1412.1895}}].

\bibitem{Donnelly:2016auv}
W.~Donnelly and L.~Freidel, \emph{{Local subsystems in gauge theory and
  gravity}}, \href{https://doi.org/10.1007/JHEP09(2016)102}{\emph{JHEP}
  {\bfseries 09} (2016) 102},
  [\href{https://arxiv.org/abs/1601.04744}{{\ttfamily 1601.04744}}].

\bibitem{Geiller:2017xad}
M.~Geiller, \emph{{Edge modes and corner ambiguities in 3d Chern-Simons theory
  and gravity}},  \href{https://arxiv.org/abs/1703.04748}{{\ttfamily
  1703.04748}}.

\bibitem{Oblak:2017ect}
B.~Oblak, \emph{{Berry Phases on Virasoro Orbits}},
  \href{https://arxiv.org/abs/1703.06142}{{\ttfamily 1703.06142}}.

\bibitem{Weinberg:2003sw}
S.~Weinberg, \emph{{Adiabatic modes in cosmology}},
  \href{https://doi.org/10.1103/PhysRevD.67.123504}{\emph{Phys. Rev.}
  {\bfseries D67} (2003) 123504},
  [\href{https://arxiv.org/abs/astro-ph/0302326}{{\ttfamily
  astro-ph/0302326}}].

\bibitem{Hinterbichler:2013dpa}
K.~Hinterbichler, L.~Hui and J.~Khoury, \emph{{An Infinite Set of Ward
  Identities for Adiabatic Modes in Cosmology}},
  \href{https://doi.org/10.1088/1475-7516/2014/01/039}{\emph{JCAP} {\bfseries
  1401} (2014) 039}, [\href{https://arxiv.org/abs/1304.5527}{{\ttfamily
  1304.5527}}].

\bibitem{Narasimhan:1979kf}
M.~S. Narasimhan and T.~R. Ramadas, \emph{{GEOMETRY OF SU(2) GAUGE FIELDS}},
  \href{https://doi.org/10.1007/BF01221361}{\emph{Commun. Math. Phys.}
  {\bfseries 67} (1979) 121--136}.

\bibitem{Babelon:1980uj}
O.~Babelon and C.~M. Viallet, \emph{{On the Riemannian Geometry of the
  Configuration Space of Gauge Theories}},
  \href{https://doi.org/10.1007/BF01208272}{\emph{Commun. Math. Phys.}
  {\bfseries 81} (1981) 515}.

\bibitem{Grabiak:1986si}
M.~Grabiak, B.~Muller and W.~Greiner, \emph{{Geometrical Properties of Gauge
  Theories}}, \href{https://doi.org/10.1016/0003-4916(86)90025-4}{\emph{Annals
  Phys.} {\bfseries 172} (1986) 213--242}.

\bibitem{Fuchs:1994zv}
J.~Fuchs, M.~G. Schmidt and C.~Schweigert, \emph{{On the configuration space of
  gauge theories}},
  \href{https://doi.org/10.1016/0550-3213(94)90128-7}{\emph{Nucl. Phys.}
  {\bfseries B426} (1994) 107--128},
  [\href{https://arxiv.org/abs/hep-th/9404059}{{\ttfamily hep-th/9404059}}].

\bibitem{Orland:1996hm}
P.~Orland, \emph{{The Metric on the space of Yang-Mills configurations}},
  \href{https://arxiv.org/abs/hep-th/9607134}{{\ttfamily hep-th/9607134}}.

\bibitem{salamon1998notes}
D.~Salamon, \emph{Notes on flat connections and the loop group},
  {\emph{Preprint, University of Warwick} (1998) }.

\bibitem{Hawking:2016msc}
S.~W. Hawking, M.~J. Perry and A.~Strominger, \emph{{Soft Hair on Black
  Holes}}, \href{https://doi.org/10.1103/PhysRevLett.116.231301}{\emph{Phys.
  Rev. Lett.} {\bfseries 116} (2016) 231301},
  [\href{https://arxiv.org/abs/1601.00921}{{\ttfamily 1601.00921}}].

\bibitem{Afshar:2016uax}
H.~Afshar, D.~Grumiller and M.~M. Sheikh-Jabbari, \emph{{Black Hole Horizon
  Fluffs: Near Horizon Soft Hairs as Microstates of Three Dimensional Black
  Holes}},  \href{https://arxiv.org/abs/1607.00009}{{\ttfamily 1607.00009}}.

\bibitem{Bousso:2017dny}
R.~Bousso and M.~Porrati, \emph{{Soft Hair as a Soft Wig}},
  \href{https://arxiv.org/abs/1706.00436}{{\ttfamily 1706.00436}}.

\bibitem{Bousso:2017rsx}
R.~Bousso and M.~Porrati, \emph{{Observable Supertranslations}},
  \href{https://arxiv.org/abs/1706.09280}{{\ttfamily 1706.09280}}.

\bibitem{KowVa}
O.~Kowalski and L.~Vanhecke, \emph{{Riemannian manifolds with homogeneous
  geodesics}}, {\emph{Boll. Un. Mat. Ital. B} {\bfseries 07} (1991) 189–246}.

\bibitem{Lee:1990nz}
J.~Lee and R.~M. Wald, \emph{{Local symmetries and constraints}},
  \href{https://doi.org/10.1063/1.528801}{\emph{J. Math. Phys.} {\bfseries 31}
  (1990) 725--743}.

\bibitem{Wald:1993nt}
R.~M. Wald, \emph{{Black hole entropy is the Noether charge}},
  \href{https://doi.org/10.1103/PhysRevD.48.R3427}{\emph{Phys. Rev.} {\bfseries
  D48} (1993) R3427--R3431},
  [\href{https://arxiv.org/abs/gr-qc/9307038}{{\ttfamily gr-qc/9307038}}].

\bibitem{Iyer:1994ys}
V.~Iyer and R.~M. Wald, \emph{{Some properties of Noether charge and a proposal
  for dynamical black hole entropy}},
  \href{https://doi.org/10.1103/PhysRevD.50.846}{\emph{Phys. Rev.} {\bfseries
  D50} (1994) 846--864}, [\href{https://arxiv.org/abs/gr-qc/9403028}{{\ttfamily
  gr-qc/9403028}}].

\bibitem{Barnich:2001jy}
G.~Barnich and F.~Brandt, \emph{{Covariant theory of asymptotic symmetries,
  conservation laws and central charges}},
  \href{https://doi.org/10.1016/S0550-3213(02)00251-1}{\emph{Nucl. Phys.}
  {\bfseries B633} (2002) 3--82},
  [\href{https://arxiv.org/abs/hep-th/0111246}{{\ttfamily hep-th/0111246}}].

\bibitem{Seraj:2016cym}
A.~Seraj, \emph{{Conserved charges, surface degrees of freedom, and black hole
  entropy}}, Ph.D. thesis, IPM, Tehran, 2016.
\newblock \href{https://arxiv.org/abs/1603.02442}{{\ttfamily 1603.02442}}.

\bibitem{Woodhouse:1980pa}
N.~Woodhouse, \emph{{Geometric Quantization}}.
\newblock Oxford University Press, 1980.

\end{thebibliography}\endgroup
\end{document}